\documentclass[sigconf, nonacm]{acmart}

\usepackage{xcolor}
\usepackage{listings}
\usepackage{multirow}
\usepackage[noend]{algorithmic}
\def\ojoin{\setbox0=\hbox{$\bowtie$}%
	\rule[0.15ex]{.22em}{.6pt}\llap{\rule[0.9ex]{.22em}{.6pt}}}

\def\fullouterjoin{\mathbin{\ojoin\mkern-5.5mu\bowtie\mkern-5.5mu\ojoin}}

\begin{document}
	%\title{A Unified Transferable Model for Learning-based DBMS [Vision]}
	\title{A Unified Transferable Model for ML-Enhanced DBMS}
	
	%%
	%% The "author" command and its associated commands are used to define the authors and their affiliations.
	\author{Ziniu Wu$^{2}$, Pei Yu$^{\#,1,4}$, Peilun Yang$^{\#,1,3}$, Rong Zhu$^1$, Yuxing Han$^1$, Yaliang Li$^1$, Defu Lian$^4$, Kai Zeng$^1$,  Jingren Zhou$^1$}
	\affiliation{%
		\smallskip
		\institution{\textit{$^{1}$Alibaba Group,
				$^{2}$Massachusetts Institute of Technology,
				$^{3}$University of Technology Sydney, 
				$^{4}$University of Science and Technology of China} \smallskip}
		\state{\textsf{$^1\{$yangpeilun.ypl, yupei.yu, red.zr, yuxing.hyx, yaliang.li, zengkai.zk, jingren.zhou$\}$@alibaba-inc.com}, \break \textsf{$^2$ziniuw@mit.edu}}, \textsf{$^4$liandefu@ustc.edu.cn}}

	\begin{abstract}
		
		Recently, the database management system (DBMS) community has witnessed the power of machine learning (ML) solutions for DBMS tasks. Despite their promising performance, these existing solutions can hardly be considered satisfactory. 
		First, these ML-based methods in DBMS are not effective enough because they are optimized on each specific task, and cannot explore or understand the intrinsic connections between tasks.
		Second, the training process has serious limitations that hinder their practicality, because they need to retrain the entire model from scratch for a new DB. Moreover, for each retraining, they require an excessive amount of training data, which is very expensive to acquire and unavailable for a new DB.
		We propose to explore the transferabilities of the ML methods both across tasks and across DBs to tackle these fundamental drawbacks.
		
		In this paper, we propose a unified model MTMLF that uses a multi-task training procedure to capture the transferable knowledge across tasks and a \emph{pre-train fine-tune} procedure to distill the transferable meta knowledge across DBs. 
		We believe this paradigm is more suitable for cloud DB service, and has the potential to revolutionize the way how ML is used in DBMS. 
		Furthermore, to demonstrate the predicting power and viability of MTMLF, we provide a concrete and very promising case study on query optimization tasks. 
		Last but not least, we discuss several concrete research opportunities along this line of work.

	\end{abstract}
	
\maketitle
	
	%\begingroup
	%\renewcommand\thefootnote{}\footnote{\noindent
	%	$\#$ The second and third authors contribute equally to this paper. 
	%}
	%\addtocounter{footnote}{-1}
	%\endgroup

	%\section{Introduction}
	\section{Introduction}

Database management system (DBMS)
is the cornerstone of a broad range of applications such as big data platforms, cloud computing, internet of things, and artificial intelligence. 
Designing and tuning DBMS involves a series of complicated tasks ranging from physical design, configuration tuning, to query optimization and execution scheduling, which all require intensive expertise.
With the growth of data volume and complexity, it becomes increasingly difficult to maintain DBMS purely using human efforts.
%to optimize its performance, e.g., query latency and throughput, is a longstanding difficult work. 
%Traditional methods rely on empirical methodologies and specifications from DB experts. 
%They are often expensive and can not generalize to different settings.

Recently, the prosperity of machine learning (ML), especially deep learning, helps to resolve a large number of DBMS challenges. ML techniques enable automatic, fine-grained, and more accurate characterization of the problem space and benefit a variety of tasks in DBMS.
Specifically, unsupervised ML techniques can model the data distribution for cardinality estimation (CardEst)~\cite{zhu2020flat, deepDB, wu2020bayescard, naru, NeuroCard} and indexing~\cite{kraska2018case, ding2020tsunami, ding2020alex, nathan2020learning}; supervised ML models can replace the cost estimator (CostEst)~\cite{sun2019end, marcus2019plan, siddiqui2020cost} and execution scheduler~\cite{marcus2016wisedb, sheng2019scheduling}; and reinforcement learning methods solve decision making problems such as configuration tuning~\cite{zhang2019end, li2019qtune, basu2016regularized} and join order selection (JoinSel)~\cite{marcus2019neo, marcus2018deep, guo2020research, yu2020reinforcement, ortiz2018learning}. 
%Enough evidence in the literature works has validated the effectiveness of ML-based solutions in DBMS.

\smallskip
%\vspace{0.2em}
\noindent\underline{\textbf{Motivation:}} 
Despite these ML methods' promising results on each individual task, the existing ML techniques in DBMS do not explore the following \textit{transferabilities} and can lead to impractical solutions and/or ineffective models. 

(1) \textit{Transferability across databases}: Existing ML methods for DBMS only focus on learning the \textit{database-specific knowledge} and ignore the \textit{database-agnostic meta knowledge} that can be transferred to new DBs. 
%They generally require an excessive and impractical amount of data (such as executed queries) to train, which might be very expensive to acquire, especially for a new database~\cite{ma2020active}.
Therefore, they need to retrain the entire model from scratch for a new DB, and generally require an excessive and impractical amount of data, such as executed queries and logs, for each retraining, which is very expensive to acquire especially for a new DB~\cite{ma2020active}
(referred to as the notorious ``cold-start'' problem).
Fortunately, some \textit{meta knowledge} can be distilled and shared across DBs to mitigate this problem. 
This knowledge (such as expert experience and heuristics in the physical join implementation and access path selection) is independent of each specific DB.
For example, the query optimizer usually chooses an index scan for high-selectivity predicates and a sequential scan for low-selectivity ones; and the hash join is usually more memory-intensive than nest loop join and merge join.
This knowledge should be distilled and shared across various databases to avoid the redundant learning process and mitigate the ``cold-start'' problem. 

(2) \textit{Transferability across tasks}: Existing ML approaches are only optimized on individual DBMS tasks and neglect the \textit{task-shared knowledge}, leading to inefficient use of data and ineffective model. 
Since all these approaches are fundamentally based on understanding the data distributions and query workload representation, the shared knowledge can be used to reduce model redundancy and improve data efficiency across tasks. 
More importantly, it can enhance the model effectiveness because these tasks are inter-dependent and this knowledge can capture the inherent interactions.
For example, the purpose of CardEst model is to help generate better query plans. 
However, different estimations have different impacts on the quality of the generated plan, which is also determined by the plan enumeration method and the cost model. 
Sometimes a series of bad estimations will not lead to a worse plan, but a small estimation error of a critical sub-query can have catastrophic outcomes. 
Thus, a CardEst model trained without considering other tasks can not effectively generate better query plans~\cite{negi2021flow}.

Inspired by the recent success of the pre-trained models (e.g., BERT~\cite{devlin2018bert} and GPT-3~\cite{brown2020language}) in NLP domain, we advocate for the next generation of ML-based methods for DBMS to explore and exploit the aforementioned \textit{transferabilities} in a unified framework. 
Within this framework, the knowledge can be distilled and shared across tasks to mutually benefit all, and the meta knowledge can be reused for new DBs. 
Specifically, we propose a meta-learning paradigm that pre-trains a model on various DBs to condense the \textit{database-agnostic meta knowledge} and fine-tunes this model to fit a new DB with a small number of training examples. 
Moreover,  we propose a multi-task training procedure that simultaneously trains the model on all DBMS tasks to extract the \textit{task-shared knowledge}.

%The pre-train models (e.g., BERT~\cite{} and GPT-3~\cite{}) build deeply stacked transformer networks~\cite{} to solve a similar problem in NLP domain, where the multi-tasking models can achieve near-human accuracy on various language understanding and question answering tasks. 
%More importantly, these models can quickly adapt to a new dataset with a few training examples (few-shot).
%Inspired by their success, the DB community has designed multi-task~\cite{} or meta-learning pre-train models~\cite{} to solve relational table understanding and data cleaning problems with very promising results. 
%Their success suggests that a similar solution might be viable for DBMS tasks because the solutions to the aforementioned problems are also essentially trying to understand the data and use the data representation to solve each specific task. 

\smallskip
\noindent \underline{\textbf{Contribution:}} 
We identify the \textit{transferable} and \textit{non-transferable} knowledge that ML models try to understand and use to solve the DBMS tasks. 
Based on the \textit{transferability} across DBs, we classify the knowledge into \textit{database-agnostic meta knowledge} and \textit{database-specific knowledge}. 
Based on the \textit{transferability} across tasks, we classify the knowledge into \textit{task-shared knowledge} and \textit{task-specific knowledge}. 

Thereafter, we propose the multi-task meta-learning framework (MTMLF) with three modules: 
(1) a featurization and encoding module to characterize the \textit{database-specific knowledge} such as the data distributions in each DB, 
(2) a shared representation module to extract \textit{task-shared knowledge} that would benefit all DBMS tasks, and 
(3) a task-specific module to tackle each task (such as CardEst, CostEst, JoinSel, indexing, and configuration tuning) and learn the \textit{task-specific knowledge}.
Furthermore, the architecture of MTMLF naturally enables a \emph{pre-train fine-tune} meta learning paradigm to distill the \textit{database-agnostic meta knowledge}. 

In order to demonstrate the viability of the envisioned MTMLF, we provide a case study \textit{MTMLF-QO} for query optimization tasks, including CardEst, CostEst, and JoinSel. Thanks to the multi-task joint learning, the \textit{MTMLF-QO} on a single DB outperforms the previous state-of-the-art (SOTA) method on CardEst/CostEst tasks, and yields near-optimal results in JoinSel task.
When trained on multiple DBs with the proposed meta-learning algorithms, \textit{MTMLF-QO} has ability to distill the \textit{meta knowledge} that can be transferred on new DBs.
%Based on this promising result, we further propose a blueprint of the meta-learning algorithm to equip the \textit{MTF-QO} with the ability to distill the  \textit{meta knowledge} that can benefit various DBs. 
%Then, we conceptually reason the effectiveness of this meta-learning algorithm and enumerate the future steps to take to practically realize this method.

%At last, we point out and elaborate a series of interesting research directions along this line of work.

The contributions of this paper are summarized as follows:

$\bullet$ We identify and classify the knowledge that ML models in DBMS essentially trying to comprehend (Section~\ref{sect2}). 

$\bullet$ We propose the MTMLF, a unified transferable model for all DBs and all tasks in DBMS (Section~\ref{sect2}). 

$\bullet$ We design a concrete model MTMLF-QO to showcase that MTMLF's viability for query optimization (Section~\ref{sect3}).
%and propose a meta-learning algorithm (in Section~\ref{sect4}) to showcase that MTMLF can deliver superior performance in query optimization tasks.

$\bullet$ We conduct experiments to demonstrate MTMLF-QO's superior performance and effectiveness of multi-task learning and multi-DB meta-learning (Section~\ref{sect4}).

$\bullet$ We point out several concrete future research directions along this line of work (Section~\ref{sect5}).

	%\section{Multi-tasking Framework}
	\section{Multi-tasking Meta-learning \break Framework}
\label{sect2}

In this section, we classify the knowledge that ML methods in DBMS trying to comprehend, from the data and task dimensions in Section~\ref{sect2.1}. 
Based on this classification, we design the multi-tasking meta-learning framework (MTMLF) to explicitly learn each type of knowledge in Section~\ref{sect2.2}. 
At last, we provide the workflow about the MTMLF with the potential to revolutionize the way how ML methods are used for DBMS in Section~\ref{sect2.3}.
%In this section, we propose a multi-tasking meta-learning framework (MTMLF), a unified framework for all tasks in DBMS, as shown in Figure~\ref{framework}. 
%We believe that it represents the next generation of ML-based method in DBMS and has the potential to revitalize the existing ones by exploring and exploiting the \textit{transferabilities} both across tasks and across DBs.
%In order to effectively learn these \textit{transferabilities}, the MTMLF should contain components to learn the following three types of information and knowledge.

%\vspace{-1em}
\subsection{Knowledge classification}
\label{sect2.1}
\begin{figure}[t]
	\centering
	\includegraphics[width=8.5cm]{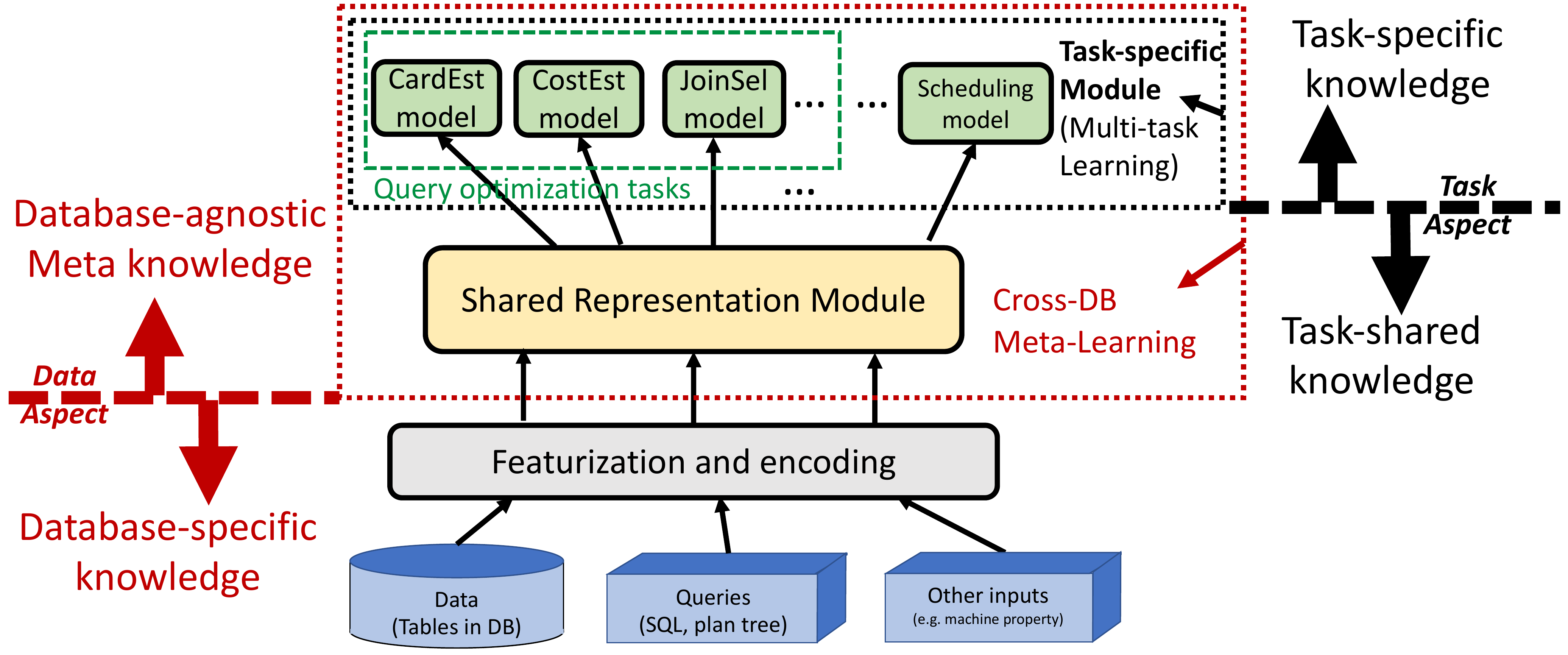}
	\vspace{-2.3em}
	\caption{Framework and knowledge overview}
	\vspace{-1em}
	\label{framework}
\end{figure}

The ML solutions for DBMS are fundamentally based on extracting knowledge from the DB and apply it to various tasks. 
We can classify this knowledge from two aspects as shown in Figure~\ref{framework}.

From data aspect based on whether the knowledge is transferable across DBs, it can be classified into \textit{database-specific} and \textit{database-agnostic meta} knowledge:

\vspace{0.2em}
\noindent \textbf{Database-specific knowledge} refers to the knowledge that is unique and can hardly benefit other DBs. Specifically, it includes the data distributions, the join schema (i.e. the fact/dimension tables and their join relationship), and the query workload in a DB. 
%the MTMLF should explicitly separate the knowledge that is not transferable from the rest. 
%Specifically, the data distributions of every single table and the query workload in a specific DB are unique and can hardly benefit from the knowledge transferred from other DBs. 
%Therefore, the MTMLF trains a \textit{featurization and encoding module} for each DB from scratch to process the non-transferable database-specific information.

\vspace{0.2em}
\noindent \textbf{Database-agnostic meta knowledge} refers to the knowledge that should be distilled and shared across various DBs.
In a high level, this knowledge is independent of the data distributions and query workloads in specific DBs, such as the expert experience and heuristics about the physical join implementation and access path selection.
E.g., to implement a hash join for foreign key join, the dimension table is usually the build side and the fact table is the probe side. 
Furthermore, in a distributed setting, the hash join is usually implemented using broadcast join where the dimension table is broadcasted.
This type of meta knowledge should be shared across DBs to avoid redundant learning process for each new DB.

From task aspect based on the knowledge's transferability across tasks, it can be classified as \textit{task-shared} and \textit{task-specific} knowledge:

\vspace{0.2em}
\noindent \textbf{Task-shared knowledge} refers to the data and query representation that can benefit all tasks in DBMS. 
The existing ML approaches to all DBMS tasks are fundamentally based on understanding the underlying data distributions and query workload representation.  
Therefore, these tasks are inter-dependent and the shared knowledge can capture their inherent interactions to enhance the model effectiveness.
For instance, the index recommender analyzes the data and query workload to recommend an index that can improve a large portion of join and scan cases encountered by the query optimizer.
Conversely, the query optimizer, which plans the execution essentially based on understanding the distributions, can generate more efficient query plans by considering the learned index. 

\vspace{0.2em}
\noindent \textbf{Task-specific knowledge} will be used to tackle each specific task based on the shared data and query workload representation.
For example, having access to the shared representation, a JoinSel model still needs to explicitly solve an NP-hard problem~\cite{howgoodare}. 
Specifically, after deriving estimated cardinalities and the cost of different operations from the shared knowledge, the JoinSel model will design specific features to solve an  NP-hard combinatorial optimization problem and plan an optimal join order.
This step is hardly beneficial to other tasks and too complex to be shared with. 

Existing ML approaches in DBMS are not effective or practical mainly because they did not explore or learn the \textit{database-agnostic meta knowledge} and \textit{task-specific knowledge}. Therefore, we propose the MTMLF to explicitly learn these two types of knowledge. 
%We believe that it represents the next generation of ML-based methods for DBMS and has the potential to revolutionalize the existing ones by exploring and exploiting the \textit{transferabilities} both across tasks and across DBs. 

%Then, it creates a \textit{task-specific module} for each specific DBMS tasks
%The MTMLF constructs a \textit{shared representation module} to extract share information of the data distributions and query workload representation that can benefit all tasks. Then, it creates a \textit{task-specific module} for each specific DBMS tasks. Compared to existing works which separately train a model for each task, the multi-task joint training of MTMLF has two advantages. First, the MTMLF is more lightweight and data-efficient to train. Second, it will be more effective because these tasks are inter-independent and the shared information can capture the inherent dependency. For example, the purpose of CardEst method is to help generate better query plans. However, different estimations have different impacts on the quality of the generated plan, which is determined by the plan enumeration method and cost model. Sometimes a series of bad estimations will not lead to a worse plan, but sometimes a small estimation error can have catastrophically outcomes~\cite{flowloss}. Thus, a CardEst model trained without considering other tasks can not effectively generate better query plans~\cite{flowloss}. 

\subsection{Framework overview}
\label{sect2.2}
We design the MTMLF with the following architecture to explicitly capture the aforementioned four types of knowledge (shown in Figure~\ref{framework}). 
First, the MTMLF uses a \textit{featurization and encoding module} for each DB to process the non-transferable \textit{database-specific knowledge}.
Second, it constructs a \textit{shared representation module} to learn the data distributions and query workload representation that can benefit all DBMS tasks. 
This module will be trained jointly on all tasks in order to extract the \textit{task-shared knowledge} and improve model effectiveness on each task. 
%In addition, joint training can help improve data efficiency and reduce model redundancy.
Third, it creates a \textit{task-specific module} with each sub-module corresponding to one DBMS task and understands the \textit{task-specific knowledge}.
Fourth, the architecture design of MTMLF enables an effective meta-learning procedure to distill the \textit{database-agnostic meta knowledge}.
All database-specific information is pushed to the \textit{featurization and encoding module}.
The remaining modules are devoted to understanding the \textit{database-agnostic knowledge}.
Therefore, we hypothesize that the \textit{task-specific} and \textit{shared representation modules} can benefit significantly from \textit{pre-train fine-tune} paradigm. I.e. we can pre-train these two modules of MTMLF using data from various DBs; thereafter, when deployed on a new DB, the pre-trained model only needs a small number of training examples to fine-tune for the best accuracy.

%\smallskip
%\vspace{0.35em}
\noindent\underline{\textbf{Featurization and encoding module}} adaptively applies feature engineering to three types of input: the data tables, the executed query workload, and additional information such as join schema and physical machine properties. 
(1) This module will take each data table in the DB as input and output its encoded distribution.
(2) This module can directly apply the existing procedures~\cite{sun2019end, marcus2019neo} to featurize each query in the workload. %, including both the raw SQL queries and the execution query plans.
(3) Some tasks may take additional information into consideration. For example, a resource allocation and execution scheduling model might need to know the memory size, buffer size, CPU usages, etc. This module can also featurize these inputs accordingly.

%\smallskip
%\vspace{0.35em}
\noindent\underline{\textbf{Shared representation module}} takes the featurized and encoded inputs, models their interactions, and outputs a shared representation that could benefit all tasks/DBs. 
For example, many tasks (e.g. CardEst, JoinSel) must understand the data distribution of the join on multiple tables. 
This module can learn such distribution by analyzing the cardinality of executed join queries and combining the single table distributions. 
Inspired by recent advance using pre-train models for NLP~\cite{brown2020language, devlin2018bert}, data cleaning~\cite{tang2020relational} and relational table understanding~\cite{herzig2020tapas, deng2020turl}, we advocate for implementing the shared representation module with transformer~\cite{transformer}, which is demonstrated to be very powerful in modeling interactions, extracting effective representations, and easy for \emph{pre-train fine-tune} procedure. 

%\vspace{0.35em}
%\smallskip
\noindent\underline{\textbf{Task-specific module}} contains a series of models corresponding to all DBMS tasks, some of which may contain many sub-tasks (for example, the query optimization task consists of CardEst, CostEst, JoinSel sub-tasks).
Each model takes the shared representation from the previous module and returns the desired outputs for its corresponding DBMS task.
This module learns the \textit{task-specific knowledge}, which can also benefit various DBs through meta-learning.
%Since these tasks are not independent, some models in the task-specific module can also take the outputs from other models as additional input. For example, the join order selection task is fundamentally based on CardEst and CostEst, as shown in dashed lines of Figure~\ref{framework}.

\subsection{Workflow overview}
\label{sect2.3}
The MTMLF has the potential to revolutionize the way how ML is used in DBMS. 
It is more suitable in the form of cloud service, which can significantly reduce the time and complexity of adopting ML-powered DBMS components (such as DB auto-tuner and learned query optimizer) on users' DBs, and boost the wide applications of these ML components.
We provide the details from the service provider and the users sides.
%By deploying the MTMLF into a cloud service, the service provider can offer users the trained MTMLF for their DBs. 
%Meanwhile, the provider can collect the executed queries from the users to further optimize the MTMLF and provide better service. The details will be given on the provider and client sides.
%The details will be given from training and inference aspects.

\vspace{0.2em}
\noindent\underline{\textbf{Service provider side:}} 
The cloud service provider will train the MTMLF on multiple users' DBs and provide the \textit{shared representation} and \textit{task-specific} modules as part of the DB service to the users. In this way, the provider can leverage its advantages: 
1) it has access to various users' DBs either through anonymous access or federated learning~\cite{konevcny2015federated} to protect the data privacy; %It will protect data privacy from users, either through data anonymization or federated learning. 
2) it has powerful computation resources to train large models;
and 3) it has abundant time because the training process is offline.
Thus, the pre-trained MTMLF can fully exploit these advanatges to distill the \textit{database-agnostic meta knowledge} beneficial to all users' DBs.
%The cloud service provider has access to a large amount of DB data gathered from users side, computing resources and large storage. It will protect data privacy from clients, either through data anonymization or federated learning. 

%The existing approaches of using ML for DBMS generally require an excessive amount of data to train a model from scratch for each task optimized on each DB.
%On the contrary, the training procedure of the MTMLF contains two phases.
%Specifically, a \textit{featurization and encoding module} is trained and fixed for each DB. Then, the remaining two modules are optimized by joint training on all tasks and all DBs. The trained model will be provided to the clients.

Furthermore, the service provider can periodically collect useful information from the users side in the form of anonymous training data or gradients of model parameters (in federated learning). 
This information will be used continuously and asynchronously to update and optimize the pre-trained MTMLF. 
The new model will be published as service upgrades to benefit all users. 
%All users can benefit from this upgrade. 
%Therefore, this training procedure forms a circle that will continuously benefit the provider and the users.

\vspace{0.2em}
\noindent\underline{\textbf{Users side:}} 
The DB users will locally adjust the received pre-trained MTMLF to best fit their DBs.
%After fine-tuning the MTMLF on the users side, the MTMLF can be used to return the desired output given appropriate inputs.
This fast local training process only requires the users to (1) analyze the data tables in the user's DB to summarize the data distributions, similar to an ``ANALYZE'' operation in traditional DBMS~\cite{postgresql}, and (2) execute a small number of representative queries to fine-tune the pre-trained MTMLF. 
%The client will receive the MTMLF from the service provider to optimize its DBMS.
%The MTMLF will analyze the data tables in the client's DB and execute a few examples queries to fine-tune the pre-trained model. 
This training procedure only needs to be conducted once and all tasks are tuned to the best performance.
%the  model produce the desired outputs for each task.
%First, immediately after training, the model can construct an index structure and recommend a set of DB configurations that would potentially speed up the query execution. 
%Then, for a new incoming query, the query optimization sub-module in MTMLF will predict its optimal query execution plan. 
%At last, taking the generated plan as input, the scheduling sub-module will schedule and allocate the computing resource during actual execution.

The MTMLF model is very effective in inference. Since all tasks are trained jointly to learn the task inherent interactions, the inference of each task can effectively take others into consideration, guaranteed to make consistent decisions.
For example, the physical design wizard will only recommend indices that the query optimizer finds useful for a large portion of query workloads.

This \emph{pre-train fine-tune} paradigm can significantly reduce the management complexity.
First, the MTMLF can swiftly evolve itself as the DB changes. When the data or query workload distribution in this DB shifts, only the \textit{featurization and encoding module} of MTMLF needs to be updated without affecting the other two modules. 
Second, despite the diverse set of DBMS tasks, only a single model needs to be maintained, monitored, and regularly updated.

	%\section{MTF-QO: Query optimization case study}
	\begin{figure*}[t]
	\centering
	\includegraphics[width=17.5cm]{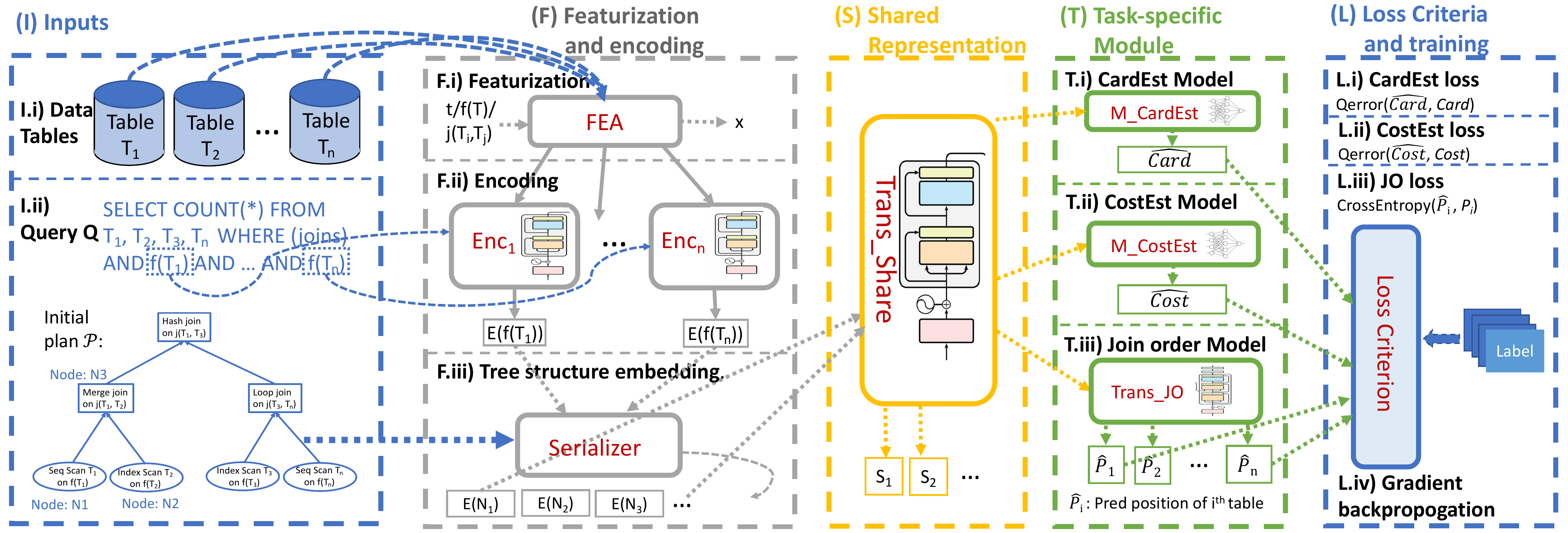}
	\vspace{-1em}
	\caption{MTMLF-QO Model for a single DB}
	\vspace{-1em}
	\label{MTF-QO}
\end{figure*}

\section{Case Study Query Optimization}
\label{sect3}

In this section, we describe the MTMLF in a concrete scenario, query optimization (QO), a key component in DBMS.  In Section~\ref{sect3.1}, we first review the relevant learning tasks in QO: CardEst, CostEst, and JoinSel tasks. 
Then in Section~\ref{sect3.2}, we present the case study model, MTMLF-QO for a single DB, which can be jointly trained on these three tasks to mutually benefit all. 
At last in Section~\ref{sect3.3}, we explain the meta-learning algorithm of MTMLF-QO for multiple DBs, which helps distill the meta-knowledge beneficial to all DBs.

\subsection{Learning tasks in query optimization}
\label{sect3.1}

The query optimizer, which takes as input a SQL query and outputs its physical execution plan, directly determines the performance of DBMS. Tuning the query optimizer is a challenging task, requiring thousands of expert-engineering hours~\cite{marcus2019neo, howgoodare}. Thus, numerous efforts have been devoted to optimizing QO using ML techniques~\cite{zhou2020database}. 
%In this section, we briefly review the learning problems in QO and their relations as follows.

Each candidate plan of a query $Q$ can be regarded as a tree, where each leaf node is a (sequential or index) scan operation on some tables and each inner node is a (merge, nested loop, or hash) join operation between multiple tables.
Following previous work~\cite{marcus2019neo}, as we focus on JoinSel, we omit other physical operations (e.g. aggregate or hash). 
%Each sub-tree of $T$ is a plan for a sub-query of $Q$. 
The QO process would enumerate several candidate plans, estimate their cardinality and cost, and select the optimal one. Next, we list the core learning tasks of QO:

$\bullet$ \emph{Cardinality estimation (CardEst)} refers to estimating the number of tuples satisfying a query before its execution. ML-based CardEst techniques try to build either unsupervised models characterizing the data distribution~\cite{deepDB} or supervised models mapping featurized queries to the cardinality~\cite{MSCN}. Recent evaluation results~\cite{zhu2020flat} have exhibited their superiority over traditional methods. 

$\bullet$ \emph{Cost estimation (CostEst)} refers to estimating the latency of a (sub-)query execution plan. ML-based CostEst methods use tree-based models (such as tree convolution~\cite{marcus2019plan} and tree-LSTM~\cite{sun2019end}) to encode a plan and map the encoding to its estimated costs.

$\bullet$ \emph{Join order selection (JoinSel)} decides the order with minimal cost to join multiple tables in the query. It is an NP-hard problem with a large search space~\cite{howgoodare}. Existing ML-based solutions attempt to effectively solve JoinSel using deep reinforcement learning techniques~\cite{marcus2019neo, marcus2018deep, guo2020research, yu2020reinforcement}. 

These core tasks in QO are interdependent. Specifically, CostEst is fundamentally based on CardEst; JoinSel requires CardEst and CostEst to evaluate the quality of the join order. It has been shown empirically that the CardEst model learned without considering the join order and cost model will not generate effective prediction~\cite{negi2021flow}. Thus, these tasks will be learned jointly in MTMLF-QO.

\subsection{Architecture}
\label{sect3.2}
As a concrete case study of the aforementioned MTMLF, the MTMLF-QO model also consists of inputs (I), featurization and encoding module (F), shared representation module (S), task-specific module (T), and loss criteria and training (L), as shown in Figure~\ref{MTF-QO}.

%\vspace{0.2em}
\smallskip
\noindent\underline{\textbf{(I) Inputs:}} The MTMLF-QO model takes two types of inputs: (I.i) the data tables $T = \{T_1, T_2, \ldots, T_n\}$ in the DB and (I.ii) the query $Q=(T_Q, j_Q, f_Q)$ where $ T_Q \subseteq T$ denotes the tables touched by $Q$, $j_Q=\{j(T_1, T_2), \ldots\}$ denotes the join predicates, and $f_Q=\{f(T_1), f(T_2),\break \ldots, f(T_n)\}$ denotes the filter predicates.
We also provide $Q$'s initial plan $\mathcal{P}$ with each node corresponding to a join or scan operation.

We modify the CardEst and CostEst tasks to let MTMLF-QO take the query $\mathcal{P}$ and estimate the cardinality and cost of the sub-plan rooted at each node of $\mathcal{P}$. 
All three tasks will be trained jointly. 

%MTF-QO will also take the data tables as input. A
%\vspace{0.2em}
\smallskip
\noindent\underline{\textbf{(F) Featurization and encoding module:}} 
We extract the useful information from each data tuple and input queries, and embed them into vectors~\cite{sun2019end, marcus2019neo} (shown in F.i). 
Specifically, we provide a value embedding for each unique column domain value in the DB to embed the tuples and the predicates $j(Q)$, $f(Q)$. We use a one-hot vector to represent each distinct table, column, and physical operation of a DB.

After the featurization, (F.ii) of this module will encode the data distribution of each table. 
For each single table $T_i$,  this module deploys a transformer encoder~\cite{transformer} ($Enc_i$), which takes the filter predicate on this table $f(T_i)$ and outputs $E(f(T_i))$ representing the distribution of $T_i$ after applying $f(T_i)$. 
%Then, the SQL query $Q$ is featurized as a sequence of join and filter predicates $E(Q)$, such as $(E(f(T_1)), E(f(T_2)), j(T_1, T_2), \ldots)$.

Thus far, we can embed each node $N_i$ of the query plan $\mathcal{P}$ as a concatenation of the one-hot vector embedding of tables touched by $N_i$, the one-hot vector of operation type, and the embedding for predicate $j(N_i)$ or encoded $E(f(N_i))$. We denote the embedding of $N_i$ as $E(N_i)$.
At last, a \textit{serializer} (F.iii) will convert the tree-structured plan into a vector $E(\mathcal{P}) = (E(N_1), E(N_2), \ldots)$ using the transformers' tree positional embedding techniques~\cite{shiv2019novel}.

%\vspace{0.2em}
\smallskip
\noindent\underline{\textbf{(S) Shared representation module:}}
After the previous module produces a sequence of embeddings $E(\mathcal{P})$ for the query, MTMLF-QO will model the interactions among elements $E(N_i)$ of $E(\mathcal{P})$ and generate a shared representation for the subsequent tasks. 
We use a transformer encoder \textit{Trans\_Share} to learn such interactions.

%The input to \textit{Trans\_Share} is the concatenation of $E(Q)$ and $E(\mathcal{P})$, i.e. $ (E(f(T_i)), E(j(T_i, T_j)), \ldots, E(N_1), E(N_2), \ldots )$, containing the information of single table distributions and join operations. 
The input $E(\mathcal{P})$ to \textit{Trans\_Share} contains the information of single table distributions and join predicates. 
\textit{Trans\_Share} will construct the multi-table join distributions and understand the cost of different physical operations on specific single and join tables. 
%We will pad $E(\mathcal{P})$  with zeros if the query plan is not presented (e.g. for the join order selection task). 
The output $(S_1, S_2, \ldots)$ of \textit{Trans\_Share} has the same length as the input, with one-to-one correspondence. 
For example, the $S_i$ corresponding to $E(N_i)$ will represent the query $\mathcal{P}$'s sub-plan rooted at node $N_i$.
%E.g., the $S_i$ corresponding to $E(j(T_i, T_j))$ will represent the join table distribution after applying the join predicate $j(T_i, T_j)$; the $S_i$ corresponding to $E(N_i)$ will be the representation of query $\mathcal{P}$'s sub-plan rooted at node $N_i$.

%\vspace{0.2em}
\smallskip
\noindent\underline{\textbf{(T) Task-specific module:}} 
Two multiple-layer perceptrons \break (MLPs), namely \textit{M\_CardEst} and \textit{M\_CostEst} can directly extract the estimated cardinality $\widehat{Card}$ and cost $\widehat{Cost}$ from the shared representation (T.i and T.ii,  respectively). However, extracting the optimal join order from this representation is much more complicated, because there exists an exponential number of possible join orders and a large amount of them might not be executable (e.g. there does not exist a join predicate between two tables). 
%Therefore, our join order selection sub-task is essentially trying to solve an NP-hard problem~\cite{IMDB} and has to guarantee the executability at the same time. 

As demonstrated in (T.iii), we formulate the JoinSel task into a sequence to sequence learning task (seq2seq) and use a transformer decoder~\cite{transformer} \textit{Trans\_JO} to generate the join order. 
For clarity of discussion, we focus on generating the left-deep join orders~\cite{howgoodare}, which can be directly flattened into an ordered sequence of tables. 
Specifically, the \textit{Trans\_JO} takes as input the shared representation $(S_1, S_2, \ldots)$, with each $S_i$ representing a single or join table. 
At each timestamp $t$, \textit{Trans\_JO} will output a value $\hat{P}_t$ representing the probability of which table should be joined at the current timestamp. 
For a DB with $n$ tables, $\hat{P}_t$ will be a multinoulli distribution vector of length $n$ with the i-th entree corresponding to the probability that the table $T_i$ is the next table to join with. 
Then, we design a novel decoding algorithm to decode a sequence of tables from the time sequence of $\hat{P}_t$ as the predicted join order, which is guaranteed to be legal and executable. 

Please note that the \textit{Trans\_JO} can also generate bushy plans with our novel decoding algorithm based on novel beam-search~\cite{boulanger2013audio, graves2012sequence}. 
We delay the discussion on the details of \textit{Trans\_JO} for bushy plans and the decoding algorithm to Section~\ref{sect4}.

%\vspace{0.2em}
\smallskip
\noindent\underline{\textbf{(L) Loss criteria and training:}} 
In order to train the models for CardEst and CostEst, we use the conventional Q-error loss~\cite{MSCN, sun2019end}, i.e., the factor between the predicted and true cardinality or cost (L.i and L.ii): $L_{card} = max(\widehat{card}/card, card/\widehat{card})$.

For the JoinSel, which can be modeled as the seq2seq task, we use the cross-entropy loss function. 
Specifically, given a ground truth optimal left-deep join order $T'_1, T'_2, \ldots, T'_m$ for a query $Q$ touching $m$ tables out of the total $n$ tables, we can embed each $T'_t$ into $P_t$, a one-hot vector of length $n$. 
At each time stamp $t$, the MTMLF-QO outputs a  probability vector  $\hat{P}_t$ and we can compute a cross entropy loss between $\hat{P}_t$ and $P_t$. 
We average the loss across all $m$ timestamps and derive the loss of join order $L_{jo} = -(\sum_{t=1}^{m} P_t  \cdot log(\hat{P}_t))/m$. This refers to the token-level loss function in NLP context~\cite{}. To further enhance the effectiveness of the model training, we design a novel sequence-level loss function to train the \textit{Trans\_JO}. We will explain the details in Section~\ref{sect5}.
%, where $\cdot$ refers to the vector inner product.

During the offline training phase of MTMLF-QO, all three tasks are trained jointly. The overall \textit{loss criterion} is defined as the weighted combination of three loss functions for three tasks as defined in equation~\ref{equ-loss}. The weights are hyper-parameters of the MTMLF-QO. 
%The weights $w_{card}$, $w_{cost}$, and $w_{jo}$ are hyper-parameters to tune.
\vspace{-2.0em}
\begin{center}
	\begin{equation}
		L_{QO} = w_{card} * L_{card} + w_{cost} * L_{cost} + w_{jo} * L_{jo}
		\label{equ-loss}
	\end{equation}
\end{center}
\vspace{-0.2em}
The gradient of this loss function will be backpropagated to update the parameters of the (S) and (T) modules only. 
%First, the featurization (F.i) and tree structure embedding (F.iii) do not need to be updated. 

For the (F) module, each single table encoder $Enc_i$ (F.ii) is trained separately with a CardEst task on a single table $T_i$. I.e. $Enc_i$ learns the data distribution of $T_i$ through predicting the cardinality of filter predicate $f(T_i)$. The details are put in the supplementary material.
%Specifically, for each table $T_i$, we analyze all filter predicates on this table from the query workload and generate a series of single-table queries $Q_i$. 
%Then, we can calculate or approximate the cardinality of $Q_i$ from on $T_i$ and use this data to train a simple model $Enc_i$ for single table CardEst. Afterward, we use the intermediate layer of learned $Enc_i$ as the embedding for the distribution of $T_i$. 

\smallskip
\noindent\underline{\textbf{Research opportunities:}} 
%The MTF-QO model opens many opportunities for future research.  First, the encoder for the single table is trained in a self-supervised fashion, which can be costly and can hardly benefit from meta-learning. 
%Thus, an efficient unsupervised approach (such as DAR~\cite{made, resMade}, SPN~\cite{SPN}, FSPN~\cite{wu2020fspn}, BN~\cite{2001SigmodGreedy}) can be used to alleviate the training cost. 
The optimal join order for a query with a large number of tables is very expensive to obtain, limiting the MTF-QO's ability to extrapolate to very complex queries. A two-phase training can potentially alleviate this problem.
I.e, an existing DBMS can be used to generate sub-optimal join orders to train a baseline MTF-QO, and then the precious data of the optimal join orders will be used to optimize this model. 

\subsection{Cross-DB meta learning for MTMLF-QO}
\label{sect3.3}
In this section, we first propose a meta-learning algorithm (MLA) for MTMLF-QO and then conceptually reason about its feasibility.

\begin{figure}[t]
	%\vspace{1em}
	\small
	\rule{\linewidth}{1pt}
	\leftline{~~~~\textbf{Algorithm 1:} \textsf{Meta-learning Algorithm for MTMLF-QO}}
	\label{MLA}
	\vspace{-1.3em}
	\begin{algorithmic}[1]
		\STATE \textbf{Input}: $n$ database $((D_1, Q_1), (D_2, Q_2), \ldots, (D_N, Q_N))$
		\STATE Initialize empty set $Train\_Data$
		\FOR{$i \in \{1, \ldots, n\}$}
		\STATE For each table $T_j$ in $D_i$, train $Enc_j$ (F.i and F.ii in Figure~\ref{MTF-QO})
		\STATE  (F) module featurizes each query in $Q_i$, and derive $E(\mathcal{P})$
		\STATE  Add $(E(\mathcal{P}), Card, Cost, P_t)$ to $Train\_Data$
		\ENDFOR
		\STATE Shuffle $Train\_Data$
		\STATE Train (S) and (T) modules with $Train\_Data$
	\end{algorithmic}
	\vspace{-1em}
	\rule{\linewidth}{1pt}
	\vspace{-3em}
\end{figure}

\smallskip
\noindent\underline{\textbf{Meta-learning algorithm:}} 
The details of MLA are shown in Algorithm~1.
Assume that MTMLF-QO has access to $n$ DBs, each with data tables $D_i$ and executed query workload $Q_i$. 
The MLA aims at enabling MTMLF-QO to predict the cardinality, cost, and join order for all $n$ DBs using a single model, and learning the \emph{database-agnostic meta knowledge}. Thus, MLA empowers MTF-QO with the ability to transfer its learned knowledge to new DBs.

First, the data tables and queries in each DB will go through the featurization module of MTF-QO. 
Then, following the training procedure described earlier, we train the single table encoder $Enc_j$ for each table $T_j$ of each DB (line 4).
Thus, the (F) module can embed each query $q \in Q_i$ with initial plan $\mathcal{P}$ (line 5) and add the embedding $E(\mathcal{P})$ and its corresponding cardinality, cost, and optimal join order to the training dataset (line 6).
After all queries in all DBs have been added, MLA will shuffle the training dataset (line 7) and train the share representation (S) and task-specific modules (T) using the aforementioned loss criteria (line 8). 

The returned MTMLF-QO trained by MLA would extrapolate to various DBs and produce accurate predictions on all of them. 
Thus, for each new DB, we can train the single table encoders ($Enc_j$ in F.ii) and the ``meta'' MTMLF-QO model only needs to be fine-tuned on a small number of example queries. 
The encoders in (F.ii) only require query cardinalities on single tables and are efficient to train.

%\smallskip
\noindent\underline{\textbf{Conceptual reasoning of MLA:}} 
MLA pushes all data-specific information to the (F) module, which can be efficiently trained for a new BD.
By shuffling the training dataset across different DBs, the MLA enforce subsequent modules of MTMLF-QO to learn the data-agnostic information, such as how the (S) module can derive the distribution on the join of multiple tables, and how the (T) module can use the shared representation to predict the cardinality, cost and join order. Without this training procedure, the (S) and (T) will likely map the embedded query to the target by brute force without truly understanding the semantics of the encoded data distribution.
We also provide a detailed example in the supplementary material on how the (S) and (T) modules can learn to construct the probability distribution on the join of multiple tables from single table distributions provided by the (F) module.

For example, for the shared representation module, the most critical and challenging part will be understanding the probability distribution on the join of multiple tables. 
Without MLA, the (S) module would require thousands of executed multi-table join queries to forcedly capture this information. Alternatively, the join tables probability distribution can be reconstructed from the single table distributions.
For example, consider two tables $A$, $B$, and their join table $O = A \fullouterjoin B$ on join predicate $A.id = B.id$. 
The probability of any filter predicate on $O$ can be derived from the distributions on $A$ and $B$ only, as shown in Equation~\ref{join}. By shuffling the training dataset across different DBs, MLA will compel the (S) module to learn this reconstruction process because otherwise a single (S) module can not extrapolate on different DBs.

\vspace{-1em}
\begin{center}
	\begin{align}
		&P_{O}(f(A) \wedge f(B)) \nonumber \\
		&= \!\!\!\!\!\!\! \sum_{id \in D(A.id)} \!\!\!\!\!\!\!  P_{A}(f(A) \wedge A.id=id) * P_{B}(f(B) \wedge B.id=id)
		\label{join}
	\end{align}
\end{center}

	\section{Join order prediction: Trans\_JO}
\label{sect4}

In this section, we present the details of the \textit{Trans\_JO} model, which takes the table representation as inputs and outputs an executable join order. Specifically, we first
explain the encoding and decoding techniques for join order trees in Section~\ref{sect4.1}. Next, we thoroughly explain the workflow of \textit{Trans\_JO} in Section~\ref{sect4.2}. At last, we present our novel beam search algorithm which guarantees to output an executable join order in Section~\ref{sect4.3}.

\subsection{Tree-to-seq and seq-to-tree conversion of query plan}
\label{sect4.1}

\begin{figure}[t]
	\centering
	%\hspace{5em}
	\includegraphics[width=0.35\textwidth]{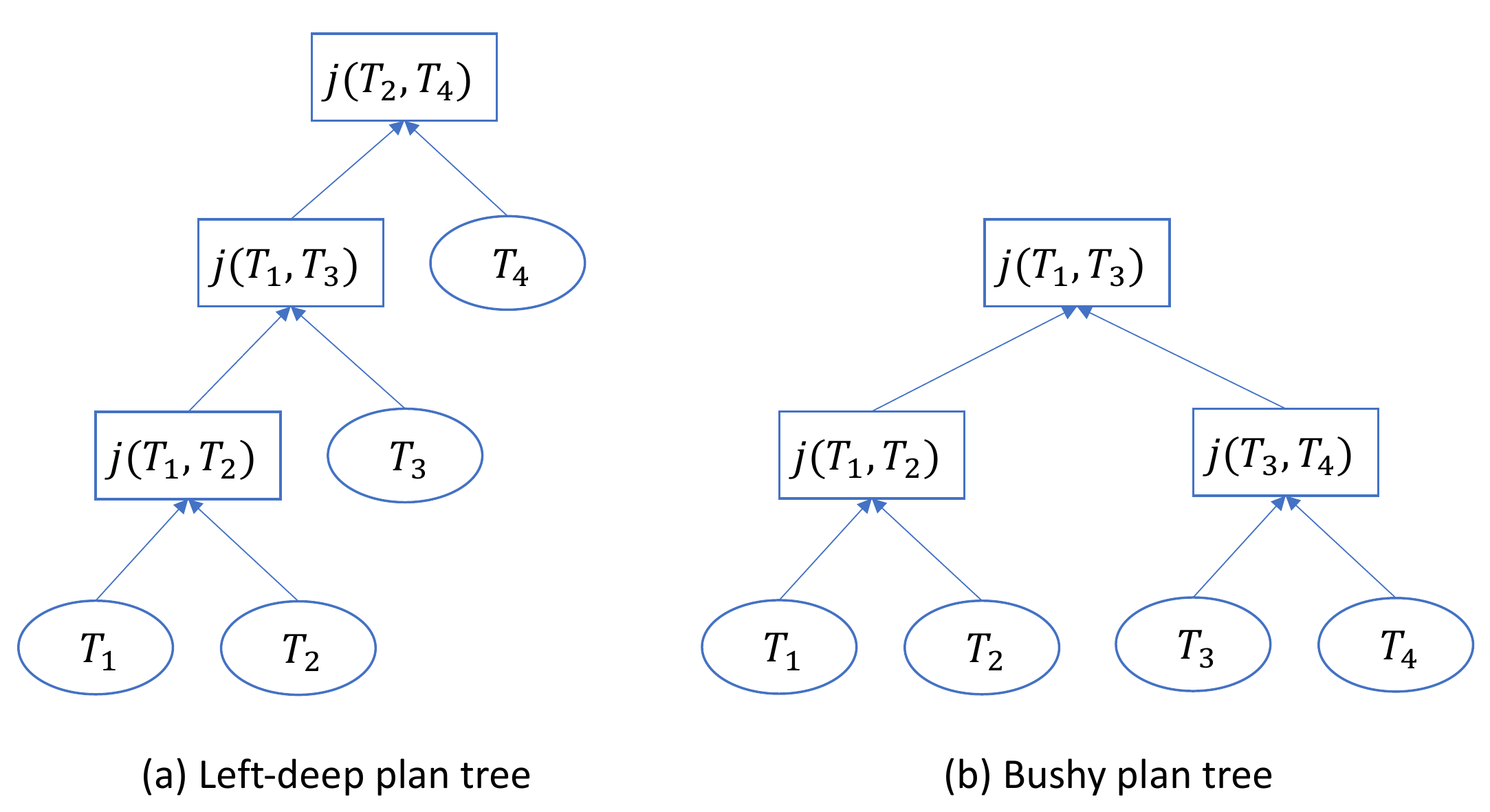}
	\vspace{-1.0em}
	\caption{Tree-structured Logic query plans}
	\vspace{-1em}
	\label{plan_tree}
\end{figure}

As described in the previous section, we formulate the join order selection task into a seq2seq learning task. Thus, it is crucial to flatten a tree-structured logical query plan into a sequence and conversely, revert a sequence into a logical query plan. We introduce our novel techniques for this tree-to-seq and seq-to-tree conversion. This method can be applied to both left-deep plans and bushy plans.

\begin{figure}[t]
	\centering
	%\hspace{5em}
	\includegraphics[width=0.35\textwidth]{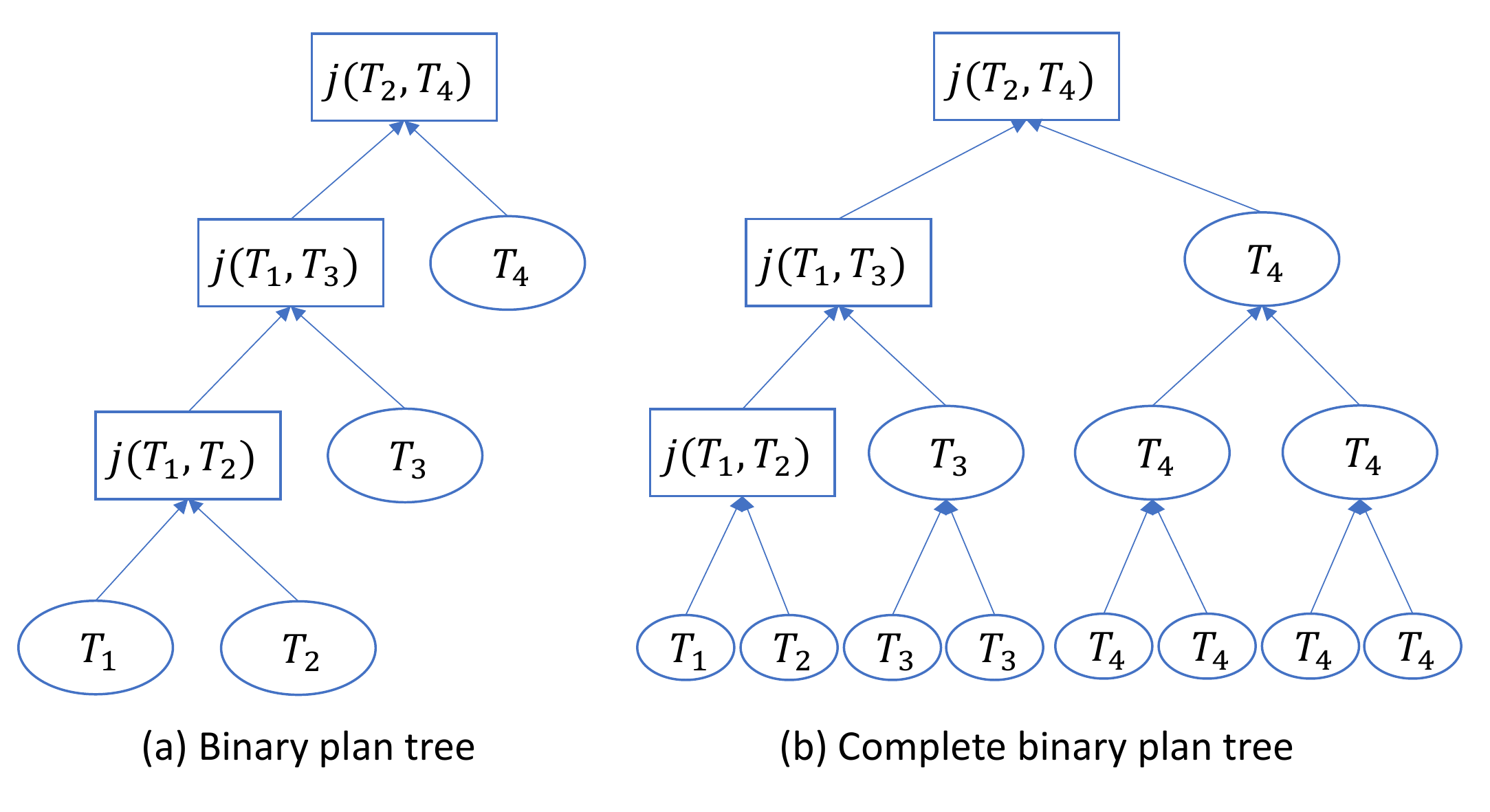}
	\vspace{-1.0em}
	\caption{Decoding techniques}
	\vspace{-1em}
	\label{decoding_tech}
\end{figure}

Since the plan tree is essentially a binary tree, we transform the plan tree into a complete binary tree.
Given a complete plan tree, we employ an embedding to record the position of each single table in the tree.
As shown in Figure \ref{decoding_tech}, the left-deep plan tree is transformed into a complete binary tree.
Leaves are noted the same as their ancestors in the tree, thus all the leaves of the sub-tree which roots at $T_3$, are marked as $T_3$. And the leaves of $T_4$ are all labeled with $T_4$.
Then, because the number of leaves of a 4-plan-tree is at most 8, we employ an 8-dimension vector as the decoding embedding to preserve the single table's position in the tree.
For each single table's embedding, the leaves which are labeled as the same single table will be set 1 and the others are all zeros.
For instance, for the left-deep plan tree (a) in the Figure \ref{plan_tree}, the decoding embeddings of $T_1, T_2, T_3, T_4$ are $[1,0,0,0,0,0,0,0], [0,1,0,0,0,0,0,0]$, $ [0,0,1,1,0,0,0,0], \break [0,0,0,0,1,1,1,1]$, respectively.
For bushy plan tree (b), the corresponding decoding embeddings are $[1,0,0,0, 0,0,0,0], \break [0,1,0,0,0,0,0,0], [0,0,1,0,0,0,0,0], [0,0,0,1,0,0,0,0]$.  $\break$
These two different plan trees can be derived from their embeddings.
Since our \textit{Tran\_JO} generates a distribution $\hat{P}_t$ at each timestamp $t$, we can compute the KL divergence between the distribution $\hat{P}_t$ and the decoding embeddings.
This loss will penalize \textit{Tran\_JO} if the prediction $\hat{P}_t$ is not the same as the ground truth position embeddings.

This method is also convenient to revert a tree.
Given all the decoding embeddings of single tables, we can note the leaves of the complete binary plan tree with the embeddings.
The leaf will be noted with the table name where this table's position embedding is 1.
For example, given an embedding $[1,0,0,0, 0,0,0,0]$ of $T_1$, the first leaf will be noted $T_1$ since only the first position in this embedding is 1.
Then, if two siblings are noted the same, their parents will be denoted the same as their label.
While if two siblings are noted differently, their parent should be a join operation $j(T_1, T_2)$.
With this novel decoding method, we can not only decode left-deep plan trees and bushy plan trees but revert a unique tree from the decoding embeddings.

\subsection{Detailed workflow of Trans\_JO}
\label{sect4.2}
At a high level, the \textit{Trans\_JO} model formulates the join order selection into a seq2seq task and uses the powerful transformer decoder~\cite{transformer} to predict the join order sequentially. We provide its detailed workflow as follows.

Assume that we have a database with $n$ tables (21 for IMDB dataset) and would like to predict the join order for a query $Q$ covering m tables $T_1, T_2,\ldots, T_m$. 
At each timestamp $t$, \textit{Trans\_JO} will output a probability distribution $\hat{P}_t$ corresponding to the position of table $T_t$ in the join order. 
%The semantic of $\hat{P}_t$ will be explained in detail in Section 2.
Similar to seq2seq task in NLP, \textit{Trans\_JO} takes two inputs at each timestamp: 
1) the outputs from the encoder layer, which are the single table representations $(S_1, S_2, ..., S_m)$ from the previously shared representation layer; 
2) the output of \textit{Trans\_JO} from previous timestamp $\hat{P}_{t-1}$. 

As described in Section~\ref{sect3.2}, given the input data tables, SQL query, and query plan, the \textit{Trans\_Share} can learn the interactions between all tables and nodes in the query plan. 
Then, the \textit{Trans\_Share} outputs a sequence of $(S_1, S_2, \ldots, S_m)$, each corresponding to a representation of single table distribution and all these distributions containing the interactions with other tables and nodes in the query plan.
Analogous to the transformer for seq2seq tasks in the NLP domain, the \textit{Trans\_Share} can be considered as an encoder, which models the interactions between tables (source language in NLP) and provides an encoding for them.
The \textit{Trans\_JO} can be considered as a decoder, which decodes the information and provides a sequence of join order for these tables (target language in NLP).

During training, the optimal join order $P_1, P_2, \ldots, P_m$ will be provided as the ground truth. A cross-entropy loss will be calculated between each pair of $P_i$ and $\hat{P}_i$ and the gradient will be backpropagated to update the parameters of  \textit{Trans\_Share} and \textit{Trans\_JO}. We also use the ``teacher forcing'' strategy~\cite{} to help our model avoid exposure bias in the seq2seq learning procedure and effectively learn the join order selection task.
%Besides single table representations, we also feed ground truth join orders of training query plans to the \textit{Tans\_JO}, which is the target sequence of the transformer decoder.
%The ground truth order $g_t$ at each timestamp is composed of the index of a single table and a one-hot embedding, which indicates what table it is. 
%The latter one is a $21$-dimension vector because there are $21$ tables in the IMDB dataset.
%At each time stamp $t$, the transformer decoder takes the previous time stamp's ground truth join order $p_{t-1}$ as input and interacts it with the memories $(m_1, m_2, ..., m_n)$. 
%And we can obtain an embedding $p_t$ which can be used to calculate the probability of which table should be joined at the current timestamp $t$.
%During training, instead of feeding ground truth join orders to \textit{Tans\_JO}, we exploit scheduled sampling strategy where \textit{Tans\_JO} takes its prediction at last timestamp $t-1$ as the input.
%This strategy helps our model avoid exposure bias problems in the sequence to sequence procedure and makes our model learn the join order selection task more effectively.
Assume that there are $n$ tables in the DB, the output distribution $\hat{P}_t$ is formulated as an n-dimension vector.
Each dimension of $\hat{P}_t$ corresponds to a specific table.
And each value on the corresponding dimension is the probability that this table should be selected.
%that the model determines as the next table to join.
At each timestamp, we choose the table whose value is the largest as the predicted table to be joined with previous tables.
%If the number of tables $m$ is less than 7, we only employ the first $m$ dimensions and the remaining dimensions will be padded with zeros.
After decoding, we can obtain the predicted join order. 
Note that our predicted join order is guaranteed to be legal and executable by using our novel beam-search algorithm which will be introduced in the following section 3.

\subsection{Join order beam search algorithm}
\label{sect4.3}

In this section, we will introduce the application of the Beam Search algorithm on \textit{Trans\_JO} and the method which can ensure the legality of output.

As described above, because we only consider the left-deep plan, \textit{Trans\_JO} can be thought of as a seq2seq task which means that the process of generating join order is equivalent to generating a sequence, i.e. the table in \textit{Trans\_JO} can be analogous to the word in machine translation. On the one hand, choosing the table with the maximum conditional probability at each generation step is easier to fall into local optimum. On the other hand, the only result generated by our model 
may not be legal which can be defined that plan with this join order can not be executed in DB, i.e. there does not exist join keys between two tables.
Thus we design a pruning strategy based on beam search which not only expands the exploration space but also ensures the legitimacy of the generated result.
So that the certain legal output we generate is more likely to approach the optimum. 

To be specific, we usually take the top $k$,  which is generally called beam width, tables as candidates for further expansion in the conditional probability distribution of the next table at each time step, and each exploration branch stops until all tables have been traversed. We finally choose the join order with the maximum probability among candidates as result. Mathematically, the number of candidates will be $k^{n-k} * k! $ if table number is $n$ and beamwidth is set to $k$ and assuming $n > k$, and we typically set the upper limit due to the excessive number. Note that a SQL query with join condition offers the information of relationship among involved tables, i.e. \textit{movie\_info.movie\_id = title.id} suggests that there is a join key between \textit{movie\_info} and \textit{title}. Thus we utilize this relationship to construct a corresponding adjacency matrix for each query. And in the process of beam search, we only choose candidates from tables having join key with current joined table in accordance with this matrix at each time step. After selection, we perform AND operation on the adjacency vector of the selected table and current joined table. Until all legal tables have been selected, we will acquire a legal candidate set to choose the best one. In that way, we ensure the final join order is executable.

	\section{Novel sequential loss criteria for join order}
\label{sect5}

In this section, we propose a novel loss function for the join order selection task, which is inspired by the works in NLP domain on sequential loss functions~\cite{ranzato2015sequence}. 

We observe that the model \textit{Trans\_JO} is trained by a token-level loss function described in Section~\ref{sect3.2} but later decoded using a sequence-level beam search algorithm as described in Section~\ref{sect4.3}. 
The token-level join order loss function is less effective because it only maximizes the conditional probability of one table in the join order but does not optimize the quality of the entire join order sequence. Therefore, we need to design a new sequence-level loss criterion to evaluate the overall join order quality as a sequence, update the model parameter accordingly, and force the model to output a more effective join order.

\smallskip
\noindent \underline{\textbf{Join order evaluation understudy:}} Inspired by the well-recognized evaluation criterion for a sequence of natural language: bilingual evaluation understudy (BLEU)~\ref{papineni2002bleu}, we first design an evaluation criterion to measure the difference between the generated join order sequence and the optimal sequence, namely Join order evaluation understudy (JOEU).

JOEU is motivated by the observation that if the partial join order of tables up to the current timestamp $t$ is not optimal, the overall join order can not be optimal regardless of the table orders after $t$. Therefore, we define $JOEU(u, u*)$ as the length of the same prefix of the two join order sequences $u$ and $u*$ divided by the sequence length. The larger JOEU of generated join order, the closer it is to the optimal join order and more likely to produce better run-time performance. 

\smallskip
\noindent \underline{\textbf{Sequence-level join order loss criterion:}}
Based on beam search and JOEU, we design a novel sequence-level loss function, as shown in Equation~\ref{seq-loss}, where $x$ is the input, $\mathcal{U}(x)$ is the candidate set of legal join order generated by beam search,
$\overline{\mathcal{U}}(x)$ is the set of illegal join orders generated by beam search and $u^{*} $ is the optimal join order (i.e. training label).

The loss function consists of three parts. 
The first part measures the negative log-likelihood of the optimal join order, which we would like to minimize so that the model is more likely to produce optimal join order.
The second part is a weighted summation of all legal join order sequences. Recall that $JOEU(u, u*)$ is a value between $0$ and $1$, with $1$ indicating the optimal join order sequence. Therefore, we want to minimize the likelihood of all join order sequences with small $JOEU(u, u*)$.
The third part penalizes illegal join orders. The penalty $\lambda$ is usually set to a large value to ensure the model would be less likely to output illegal join order. 
Intuitively, minimizing the $\mathcal{L}_{JO}$ can maximize the occurrence probability of the optimal join order and minimize the likelihood of both not optimal but legal join order and illegal join order.

\begin{center}
	\begin{align}
		\mathcal{L}_{JO}  &= - log(p(u^{*}|x)) + \!\!\! \sum_{u\in\mathcal{U}(x)} \!\! (1-JOEU(u, u*))*log(p(u|x)) \nonumber \\
		& + \lambda * log\sum_{u\in\overline{\mathcal{U}}(x)}p(u|x)
		\label{seq-loss}
	\end{align}
\end{center}

%\begin{equation}
%\mathcal{L}_{SeqNLL} = - logp(u^{*}|x) + (1-JoEU(u))*log\sum_{u\in\mathcal{U}(x)}p(u|x) 
%+ log\sum_{u\in\overline{\mathcal{U}}(x)}p(u|x)
%\end{equation}

	%\section{Cross DB meta learning}
	\section{Experimental results}
\label{sect6}
We first evaluate the performance of MTMLF-QO for a single DB on CardEst, CostEst, and JoinSel tasks and the effectiveness of multi-task joint training (Section~\ref{sect6.1}). 
Then, we provide a data generation pipeline to generate artificial DBs (Section~\ref{sect6.2}). These DBs will be used to evaluate the cross-DB ``transferability'' of MTMLF-QO trained via MLA (Section~\ref{sect6.3}).

\begin{table}[t]
	\centering
	\caption{Q-errors on the JOB workload.}
	%\vspace{-1.3em}
	\label{tab:exp:ccest}
	\scalebox{0.72}{
		\begin{tabular}{|c|c c c|c c c|}
			\hline
			\multirow{2}{*}{Method} & \multicolumn{3}{c|}{Cardinality} & \multicolumn{3}{c|}{Cost} \\
			& median & max & mean & median & max & mean \\ \hline
			PostgreSQL & 184.00 & 670,000 & 10,416 & 4.90 & 4920 & 105.00 \\
			%MySQL & 104.00 & 2,487,611 & 60,229 & 7.94 & 1943 & 173.00 \\
			%Oracle & 119.00 & 927,648 & 34,493 & 6.63 & 1274 & 55.30 \\
			Tree-LSTM & 8.78 & 696.29 & 36.83 & 4.00 & 290.35 & 15.01 \\
			\textbf{MTMLF-QO} & \textbf{4.48} & \textbf{614.45} & \textbf{28.69} & \textbf{2.10} & \textbf{37.54} & \textbf{4.20} \\ 
			MTMLF-CardEst & $5.12$ & $804.48$ & $36.66$ & $\backslash$ & $\backslash$ & $\backslash$ \\ 
			MTMLF-CostEst & $\backslash$ & $\backslash$ & $\backslash$ & $2.06$ & $61.41$ & $4.69$ \\  \hline
		\end{tabular}
	}
	%\vspace{-1.5em}
\end{table}

\begin{table}[t]
	\centering
	\caption{Execution time with different join orders.}
	%\vspace{-1.2em}
	\label{tab:exp:exectime}
	\scalebox{0.9}{
		\begin{tabular}{|c|c|c|}
			\hline
			JoinOrder & Total Time & Overall Improvement Ratio \\ \hline
			PostgreSQL & 1143.2 min & $\backslash$ \\ 
			Optimal & 209.1 min & 81.7\% \\ 
			\textbf{MTMLF-QO} & \textbf{318.3 min} & \textbf{72.2\%} \\ 
			MTMLF-JoinSel & 450.4 min & 60.6\% \\ 
			\hline
		\end{tabular}
	}
	%\vspace{-1.5em}
\end{table}

\subsection{Experiments on single DB}
\label{sect6.1}

We use the JOB benchmark of $113$ queries joining dozens of tables and having the complex ``LIKE'' predicates on the IMDB dataset containing $21$ tables with skewed distribution and strong attribute correlation~\cite{howgoodare}.
Following the prior work~\cite{sun2019end}, we generate $150$K SQL queries similar to the JOB queries as the training data. 
Then, we execute these queries in PostgreSQL~\cite{postgresql} to derive the query plans and their true cardinalities and costs. 
For the JoinSel task, we generate the optimal join order using the ECQO program~\cite{trummer2019exact}. 
Since deriving optimal join order has exponential time complexity, we can only afford to execute this program for $20$K queries out of the $150$K, which touches no more than $8$ tables. 

\smallskip
\noindent\underline{\textbf{Hyperparameters of MTMLF-QO:}} 
%\noindent\underline{\textbf{Hyperparameters and experiment environment:}} 
We use a transformer with 3 blocks and 4 headers for each $Enc_i$, the \textit{Trans\_Share} and the \textit{Trans\_JO}.
%The transformer decoder used in the \textit{Trans\_JO} employs the same number of blocks and headers as the \textit{Trans\_Share}.
We use two-layer MLPs for \textit{M\_CardEst} and \textit{M\_CostEst}. The weights $w_{card}$, $w_{cost}$, and $w_{jo}$ are all set to 1.
The Adam optimizer~\cite{kingma2014adam} with $10^{-4}$ learning rate is used to optimize the model. 
All experiments are conducted on a CentOS Server with an Intel Xeon Platinum 8163 2.50GHz 64-core CPU, 376GB DDR4 main memory, and 1TB SSD and GeForce RTX 2080 Ti GPU.

\smallskip
\noindent\underline{\textbf{Performance on CardEst and CostEst:}} In order to show the effectiveness of our MTMLF-QO model on CardEst and CostEst tasks, we compare it with a traditional DBMS PostgreSQL~\cite{postgresql}, and the previous SOTA method Tree-LSTM~\cite{sun2019end} on the JOB benchmark. 
Please note that we can not compare MTMLF-QO with other unsupervised SOTA methods~\cite{zhu2020flat, wu2020bayescard, deepDB} because they can not support ``Like'' predicates on strings~\cite{han2021cardinality}. 

We take 90\% of generated $150$K queries as the training dataset, 10\% as the validation set for hyper-parameter tuning, and JOB queries as the test set.
We use q-error as the metric to evaluate cardinality and cost estimation.
As shown in Table \ref{tab:exp:ccest}, our MTMLF-QO significantly outperforms the traditional DBMS and the previous SOTA Tree-LSTM on both CardEst and CostEst tasks.

\smallskip
\noindent\underline{\textbf{Performance on JoinSel:}} 
To evaluate the quality of the join order generated by MTMLF-QO, we use $85\%$ of the $20$K queries to train, $10\%$ of the queries to find the hyper-parameter, and the rest $5\%$ as the test set to predict the optimal join orders. Note that, we refrain from testing on the original JOB queries because MTMLF-QO only has access to queries joining no more than $8$ tables.
%and thus can not extrapolate to the JOB queries joining more than ten tables.

We compare the quality of the join order generated by MTMLF-QO against two baselines: the original PostgreSQL's query optimizer and the optimal join order produced by ECQO. 
%We set \textit{join\_collapse\_limit = 1} parameter in PostgreSQL to inject the learned and optimal join orders. 
Table~\ref{tab:exp:exectime} shows the results of query execution time using different join orders, where ``total time'' is the total running time of all $1,000$ testing queries, and ``overall improvement ratio'' refers to the improvement over the PostgreSQL divided by the PostgreSQL total time.

Based on this table, we can see that the learned join order of MTMLF-QO can significantly outperform the PostgreSQL baseline. 
%In some extreme cases, we even observe up to a hundred times speed-up for MTMLF-QO over Postgres (less than one-minute running time for MTMLF-QO  but over one hour for PostgreSQL). 
In addition, for more than $70$\% of the $1,000$ testing queries, MTMLF-QO can output the optimal join order. 
%For the rest of the predicted orders that are not optimal, we still observe a shorter running time for MTMLF-QO over PostgreSQL. 
These results indicate that MTMLF-QO can be a very effective learned query optimizer of PostgreSQL. 
We left the comparison of MTMLF-QO with other SOTA join order selection methods~\cite{marcus2019neo, marcus2018deep, guo2020research, yu2020reinforcement} as future work.

\vspace{0.2em}
%\smallskip
\noindent\underline{\textbf{Benefits of multi-task joint training:}} 
In order to demonstrate the benefits of multi-task joint training of MTMLF-QO, we conduct an ablation experiment to separately train the MTMLF-QO model for CardEst (MTMLF-CardEst), CostEst (MTMLF-CostEst), and JoinSel (MTMLF-JoinSel). According to Table~\ref{tab:exp:ccest} and Table~\ref{tab:exp:exectime}, the performance of MTMLF-JoinSel is much worse than the original MTMLF-QO, and MTMLF-CardEst and MTMLF-CostEst are slightly worse than MTMLF-QO.
This suggests that the joint training of CardEst, CostEst, and JoinSel tasks is indeed more effective than the separate training.

\subsection{Data generation pipeline}
\label{sect6.2}

Since there exists a very limited number of real-world DBs open to the research community, we have to generate artificial DBs to verify the cross-DB transferability of MTMLF-QO. Specifically, we design a data generation pipeline that can automatically generate DBs containing $6-11$ tables with varied numbers of attributes and very different distributions. We provide the details of how this pipeline generates one DB step by step as follows.

\smallskip
\noindent \underline{\textbf{S1: Generate a valid join schema $\mathcal{J}$.}} The first step of generating a DB $\mathcal{D}$ will be deciding how many tables should this DB contain and what is the join relationship between these tables. We first uniformly sample a number $n$ ranging from 6 to 11 and create $n$ empty tables $\{T_1, \cdots, T_n \}$. Then, we pick $2-3$ tables as the fact tables and the rest tables as dimension tables. W.L.O.G assume that $T_1, T_2$ are the fact tables and $T_3, \cdots, T_n$ are the dimension tables.
We create the first join relation between $T_1$'s primary key and $T_2$'s foreign key. For each of the dimension table $T_i$, we create a join relation between $T_i$ and $T_1$, $T_2$ or both. 
At last, each dimension table will connect with one or two fact tables, i.e., primary key/foreign key join relation (PK-FK). There does not exist any PK-FK join relation between the dimension tables but they can be joined through a transitive FK-FK join. For example, if both foreign keys $FK3_{T1}$ of $T_3$ and $FK4_{T1}$ of $T_4$ can join with the primary key $PK1$ of $T_1$, then $FK3_{T1}$ and $FK4_{T1}$ can form a FK-FK join relation. Next, we discuss how to fill in the content of each table $T_i$.

\smallskip
\noindent \underline{\textbf{S2: Generate attribute columns for $T_i$.}} 
We first randomly generate two numbers $r$ and $c$ to be the total number of rows (between $50K$ to $10M$) and attribute columns (between $2$ to $20$) in $T_i$. Then, we generate the attribute columns of $T_i$ using two approaches. The first approach is completely artificially-generated with varied data distribution skewness, attributes correlation, and domain size. This approach is similar to recent works~\cite{wang2020ready, wu2020bayescard}.
The second approach bootstraps from an existing real-world data table. By controlling the distribution of the number of bootstrapping rows and columns, we can create data distribution with varied skewness and correlation but the domain remains the same as the original table. In this way, the generated data mimics the real-world one. 

\smallskip
\noindent \underline{\textbf{S3: Generate join keys for $T_i$.}} We first create a column of primary key (PK) for $T_i$ (unique value from 1 to $r$). Creating the columns of foreign keys (FK) is much more complicated. First, we need to identify from the join schema $\mathcal{J}$ all the fact tables that can join with $T_i$. Then, $T_i$ creates a column $FKi_j$ of FK for each fact table $T_j$ with a domain equal to the PK domain of $T_j$. Since the join keys are correlated with the attribute columns~\cite{howgoodare}, we will make the values of $FKi_j$ correlate with attributes in $T_i$.

\subsection{Experiments on cross-DB transferrability}
\label{sect6.3}

\begin{table}[t]
	\centering
	\caption{Execution time with different join orders.}
	%\vspace{-1.2em}
	\label{tab:exp:MLA}
	\scalebox{0.8}{
		\begin{tabular}{|c|c|c|}
			\hline
			JoinOrder & Total Time & Overall Improvement Ratio \\ \hline
			PostgreSQL & 393.9 min & $\backslash$ \\ 
			%Optimal & 101.6 min & 74.3\% \\ 
			\textbf{MTMLF-QO (MLA)} & \textbf{234.1 min} & \textbf{40.6\%} \\ 
			MTMLF-QO (single) & 219.5 min & 44.3\% \\ 
			\hline
		\end{tabular}
	}
	%\vspace{-1.5em}
\end{table}

\noindent \underline{\textbf{Experiment procedures:}} 
We use the aforementioned pipeline to generate 11 DBs $\{\mathcal{D}_1, \ldots, \mathcal{D}_{11}\}$. For each DB $\mathcal{D}_i$, we create a workload $W_i$ of $20K$ join queries and execute the ECQO program~\cite{trummer2019exact} to derive its optimal join order. 

The hyper-parameters of MTMLF are the same as described in Section~4.1 of the main paper. 
The training of the MTMLF follows the MLA procedure. 
Specifically, we first generate some single-table queries for each table within each DB $\mathcal{D}_i$. Then, we learn a featurization module $F_i$ for every DB to capture all the dataset-specific knowledge such as the single table distributions. The procedure of training each $F_i$ is very efficient since the single table query can be efficiently executed in parallel or using AQP techniques.
W.L.O.G., we use the first 10 DBs $\{\mathcal{D}_1, \ldots, \mathcal{D}_{10}\}$ as the training data and learns the (S) shared representation and (T) task-specific modules for the MTMLF via MLA. 
These two modules can output effective join orders for all $10$ DBs. Thus, it must have captured the dataset-agnostic knowledge that can be transferred to a new DB.

We use $\mathcal{D}_{11}$ as testing data to verify the transferability of MTMLF.
Specifically, we connect the learned $F_{11}$ module containing all dataset-specific information of $\mathcal{D}_{11}$ with the pre-trained (S) and (T) modules.
Then, we use this MTMLF model to generate the join order of queries in $W_11$ and execute these join orders in PostgreSQL.

\smallskip
\noindent \underline{\textbf{Effectiveness of MTMLF-QO's meta-learning:}} 
From Table~\ref{tab:exp:MLA}, we observe that the MTMLF-QO trained via MLA  can generate join orders that are $40\%$ faster than the ones produced by PostgreSQL baseline on a brand new DB. As a controlled study, we directly train an MTMLF-QO on this test DB $\mathcal{D}$ from scratch (MTMLF-QO single), which is only slightly better than MTMLF-QO trained via MLA. These results suggest that MTMLF-QO can distill cross-DB meta-knowledge that is transferrable to new DBs.

	%\section{Opportunities for future research and conclusion}
	\section{Conclusions}
\label{sect7}
In this paper, we present the MTMLF, which can condense an effective shared representation to mutually benefit various tasks in DBMS and distill the ``meta-knowledge'' beneficial to all DBs. 
We also demonstrate with a very promising case study on query optimization that future research along this direction can be fruitful. 

Next, we list two concrete future research opportunities.
First, inspired by MTMLF-QO, other DBMS tasks can also be incorporated into the MTMLF framework. 
Second, a cloud DB service can greatly facilitate the \emph{pre-train fine-tune} paradigm of MTMLF. This setting motivates the research community to design a federated learning algorithm to protect the DB users' data privacy and simultaneously ensure effective training of MTMLF.

%, that are currently unavailable to the DBMS research community. 
%Thus, providing a pipeline for generating artificial databases with verisimilar data and query workload is crucial. With this pipeline, MTMLF can automatically generate training data, and learn the database-agnostic meta knowledge continuously. , which requires a large amount of DBs with data and executed queries

	%\clearpage
	
	\bibliographystyle{ACM-Reference-Format}
	\bibliography{ref}


\begin{thebibliography}{43}

%%% the Web address.

\ifx \showCODEN    \undefined \def \showCODEN     #1{\unskip}     \fi
\ifx \showDOI      \undefined \def \showDOI       #1{#1}\fi
\ifx \showISBNx    \undefined \def \showISBNx     #1{\unskip}     \fi
\ifx \showISBNxiii \undefined \def \showISBNxiii  #1{\unskip}     \fi
\ifx \showISSN     \undefined \def \showISSN      #1{\unskip}     \fi
\ifx \showLCCN     \undefined \def \showLCCN      #1{\unskip}     \fi
\ifx \shownote     \undefined \def \shownote      #1{#1}          \fi
\ifx \showarticletitle \undefined \def \showarticletitle #1{#1}   \fi
\ifx \showURL      \undefined \def \showURL       {\relax}        \fi
% invisible to TeX
\providecommand\bibfield[2]{#2}
\providecommand\bibinfo[2]{#2}
\providecommand\natexlab[1]{#1}
\providecommand\showeprint[2][]{arXiv:#2}

\bibitem[\protect\citeauthoryear{Basu, Lin, Chen, Vo, Yuan, Senellart, and
  Bressan}{Basu et~al\mbox{.}}{2016}]%
        {basu2016regularized}
\bibfield{author}{\bibinfo{person}{Debabrota Basu}, \bibinfo{person}{Qian Lin},
  \bibinfo{person}{Weidong Chen}, \bibinfo{person}{Hoang~Tam Vo},
  \bibinfo{person}{Zihong Yuan}, \bibinfo{person}{Pierre Senellart}, {and}
  \bibinfo{person}{St{\'e}phane Bressan}.} \bibinfo{year}{2016}\natexlab{}.
\newblock \showarticletitle{Regularized cost-model oblivious database tuning
  with reinforcement learning}.
\newblock In \bibinfo{booktitle}{\emph{Transactions on Large-Scale Data-and
  Knowledge-Centered Systems XXVIII}}. \bibinfo{publisher}{Springer},
  \bibinfo{pages}{96--132}.
\newblock


\bibitem[\protect\citeauthoryear{Boulanger-Lewandowski, Bengio, and
  Vincent}{Boulanger-Lewandowski et~al\mbox{.}}{2013}]%
        {boulanger2013audio}
\bibfield{author}{\bibinfo{person}{Nicolas Boulanger-Lewandowski},
  \bibinfo{person}{Yoshua Bengio}, {and} \bibinfo{person}{Pascal Vincent}.}
  \bibinfo{year}{2013}\natexlab{}.
\newblock \showarticletitle{Audio Chord Recognition with Recurrent Neural
  Networks}. In \bibinfo{booktitle}{\emph{ISMIR}}. Citeseer,
  \bibinfo{pages}{335--340}.
\newblock


\bibitem[\protect\citeauthoryear{Brown, Mann, Ryder, Subbiah, Kaplan, Dhariwal,
  Neelakantan, Shyam, Sastry, Askell, et~al\mbox{.}}{Brown
  et~al\mbox{.}}{2020}]%
        {brown2020language}
\bibfield{author}{\bibinfo{person}{Tom~B Brown}, \bibinfo{person}{Benjamin
  Mann}, \bibinfo{person}{Nick Ryder}, \bibinfo{person}{Melanie Subbiah},
  \bibinfo{person}{Jared Kaplan}, \bibinfo{person}{Prafulla Dhariwal},
  \bibinfo{person}{Arvind Neelakantan}, \bibinfo{person}{Pranav Shyam},
  \bibinfo{person}{Girish Sastry}, \bibinfo{person}{Amanda Askell},
  {et~al\mbox{.}}} \bibinfo{year}{2020}\natexlab{}.
\newblock \showarticletitle{Language models are few-shot learners}.
\newblock \bibinfo{journal}{\emph{arXiv preprint arXiv:2005.14165}}
  (\bibinfo{year}{2020}).
\newblock


\bibitem[\protect\citeauthoryear{Deng, Sun, Lees, Wu, and Yu}{Deng
  et~al\mbox{.}}{2020}]%
        {deng2020turl}
\bibfield{author}{\bibinfo{person}{Xiang Deng}, \bibinfo{person}{Huan Sun},
  \bibinfo{person}{Alyssa Lees}, \bibinfo{person}{You Wu}, {and}
  \bibinfo{person}{Cong Yu}.} \bibinfo{year}{2020}\natexlab{}.
\newblock \showarticletitle{Turl: Table understanding through representation
  learning}.
\newblock \bibinfo{journal}{\emph{arXiv preprint arXiv:2006.14806}}
  (\bibinfo{year}{2020}).
\newblock


\bibitem[\protect\citeauthoryear{Devlin, Chang, Lee, and Toutanova}{Devlin
  et~al\mbox{.}}{2018}]%
        {devlin2018bert}
\bibfield{author}{\bibinfo{person}{Jacob Devlin}, \bibinfo{person}{Ming-Wei
  Chang}, \bibinfo{person}{Kenton Lee}, {and} \bibinfo{person}{Kristina
  Toutanova}.} \bibinfo{year}{2018}\natexlab{}.
\newblock \showarticletitle{Bert: Pre-training of deep bidirectional
  transformers for language understanding}.
\newblock \bibinfo{journal}{\emph{arXiv preprint arXiv:1810.04805}}
  (\bibinfo{year}{2018}).
\newblock


\bibitem[\protect\citeauthoryear{Ding, Minhas, Yu, Wang, Do, Li, Zhang,
  Chandramouli, Gehrke, Kossmann, et~al\mbox{.}}{Ding et~al\mbox{.}}{2020a}]%
        {ding2020alex}
\bibfield{author}{\bibinfo{person}{Jialin Ding}, \bibinfo{person}{Umar~Farooq
  Minhas}, \bibinfo{person}{Jia Yu}, \bibinfo{person}{Chi Wang},
  \bibinfo{person}{Jaeyoung Do}, \bibinfo{person}{Yinan Li},
  \bibinfo{person}{Hantian Zhang}, \bibinfo{person}{Badrish Chandramouli},
  \bibinfo{person}{Johannes Gehrke}, \bibinfo{person}{Donald Kossmann},
  {et~al\mbox{.}}} \bibinfo{year}{2020}\natexlab{a}.
\newblock \showarticletitle{ALEX: an updatable adaptive learned index}. In
  \bibinfo{booktitle}{\emph{Proceedings of the 2020 ACM SIGMOD International
  Conference on Management of Data}}. \bibinfo{pages}{969--984}.
\newblock


\bibitem[\protect\citeauthoryear{Ding, Nathan, Alizadeh, and Kraska}{Ding
  et~al\mbox{.}}{2020b}]%
        {ding2020tsunami}
\bibfield{author}{\bibinfo{person}{Jialin Ding}, \bibinfo{person}{Vikram
  Nathan}, \bibinfo{person}{Mohammad Alizadeh}, {and} \bibinfo{person}{Tim
  Kraska}.} \bibinfo{year}{2020}\natexlab{b}.
\newblock \showarticletitle{Tsunami: A learned multi-dimensional index for
  correlated data and skewed workloads}.
\newblock \bibinfo{journal}{\emph{arXiv preprint arXiv:2006.13282}}
  (\bibinfo{year}{2020}).
\newblock


\bibitem[\protect\citeauthoryear{Graves}{Graves}{2012}]%
        {graves2012sequence}
\bibfield{author}{\bibinfo{person}{Alex Graves}.}
  \bibinfo{year}{2012}\natexlab{}.
\newblock \showarticletitle{Sequence transduction with recurrent neural
  networks}.
\newblock \bibinfo{journal}{\emph{arXiv preprint arXiv:1211.3711}}
  (\bibinfo{year}{2012}).
\newblock


\bibitem[\protect\citeauthoryear{Group}{Group}{2018}]%
        {postgresql}
\bibfield{author}{\bibinfo{person}{The PostgreSQL Global~Development Group}.}
  \bibinfo{year}{2018}\natexlab{}.
\newblock \bibinfo{booktitle}{\emph{Documentation PostgreSQL 10.3}}.
\newblock


\bibitem[\protect\citeauthoryear{Guo and Daudjee}{Guo and Daudjee}{2020}]%
        {guo2020research}
\bibfield{author}{\bibinfo{person}{Runsheng~Benson Guo} {and}
  \bibinfo{person}{Khuzaima Daudjee}.} \bibinfo{year}{2020}\natexlab{}.
\newblock \showarticletitle{Research challenges in deep reinforcement
  learning-based join query optimization}. In
  \bibinfo{booktitle}{\emph{Proceedings of the Third International Workshop on
  Exploiting Artificial Intelligence Techniques for Data Management}}.
  \bibinfo{pages}{1--6}.
\newblock


\bibitem[\protect\citeauthoryear{Han, Wu, Wu, Zhu, Yang, Tan, Zeng, Cong, Qin,
  Pfadler, et~al\mbox{.}}{Han et~al\mbox{.}}{2021}]%
        {han2021cardinality}
\bibfield{author}{\bibinfo{person}{Yuxing Han}, \bibinfo{person}{Ziniu Wu},
  \bibinfo{person}{Peizhi Wu}, \bibinfo{person}{Rong Zhu},
  \bibinfo{person}{Jingyi Yang}, \bibinfo{person}{Liang~Wei Tan},
  \bibinfo{person}{Kai Zeng}, \bibinfo{person}{Gao Cong},
  \bibinfo{person}{Yanzhao Qin}, \bibinfo{person}{Andreas Pfadler},
  {et~al\mbox{.}}} \bibinfo{year}{2021}\natexlab{}.
\newblock \showarticletitle{Cardinality Estimation in DBMS: A Comprehensive
  Benchmark Evaluation}.
\newblock \bibinfo{journal}{\emph{arXiv preprint arXiv:2109.05877}}
  (\bibinfo{year}{2021}).
\newblock


\bibitem[\protect\citeauthoryear{Herzig, Nowak, M{\"u}ller, Piccinno, and
  Eisenschlos}{Herzig et~al\mbox{.}}{2020}]%
        {herzig2020tapas}
\bibfield{author}{\bibinfo{person}{Jonathan Herzig},
  \bibinfo{person}{Pawe{\l}~Krzysztof Nowak}, \bibinfo{person}{Thomas
  M{\"u}ller}, \bibinfo{person}{Francesco Piccinno}, {and}
  \bibinfo{person}{Julian~Martin Eisenschlos}.}
  \bibinfo{year}{2020}\natexlab{}.
\newblock \showarticletitle{Tapas: Weakly supervised table parsing via
  pre-training}.
\newblock \bibinfo{journal}{\emph{arXiv preprint arXiv:2004.02349}}
  (\bibinfo{year}{2020}).
\newblock


\bibitem[\protect\citeauthoryear{Hilprecht, Schmidt, Kulessa, Molina, Kristian,
  and Binnig}{Hilprecht et~al\mbox{.}}{2020}]%
        {deepDB}
\bibfield{author}{\bibinfo{person}{Benjamin Hilprecht},
  \bibinfo{person}{Andreas Schmidt}, \bibinfo{person}{Moritz Kulessa},
  \bibinfo{person}{Alejandro Molina}, \bibinfo{person}{Kersting Kristian},
  {and} \bibinfo{person}{Carsten Binnig}.} \bibinfo{year}{2020}\natexlab{}.
\newblock \showarticletitle{DeepDB: Learn from Data, not from Queries!}
\newblock \bibinfo{journal}{\emph{PVLDB}} (\bibinfo{year}{2020}).
\newblock


\bibitem[\protect\citeauthoryear{Kingma and Ba}{Kingma and Ba}{2014}]%
        {kingma2014adam}
\bibfield{author}{\bibinfo{person}{Diederik~P Kingma} {and}
  \bibinfo{person}{Jimmy Ba}.} \bibinfo{year}{2014}\natexlab{}.
\newblock \showarticletitle{Adam: A method for stochastic optimization}.
\newblock \bibinfo{journal}{\emph{arXiv preprint arXiv:1412.6980}}
  (\bibinfo{year}{2014}).
\newblock


\bibitem[\protect\citeauthoryear{Kipf, Kipf, Radke, Leis, Boncz, and
  Kemper}{Kipf et~al\mbox{.}}{2019}]%
        {MSCN}
\bibfield{author}{\bibinfo{person}{Andreas Kipf}, \bibinfo{person}{Thomas
  Kipf}, \bibinfo{person}{Bernhard Radke}, \bibinfo{person}{Viktor Leis},
  \bibinfo{person}{Peter Boncz}, {and} \bibinfo{person}{Alfons Kemper}.}
  \bibinfo{year}{2019}\natexlab{}.
\newblock \showarticletitle{Learned Cardinalities: Estimating correlated joins
  with deep learning}.
\newblock \bibinfo{journal}{\emph{CIDR}} (\bibinfo{year}{2019}).
\newblock


\bibitem[\protect\citeauthoryear{Kone{\v{c}}n{\`y}, McMahan, and
  Ramage}{Kone{\v{c}}n{\`y} et~al\mbox{.}}{2015}]%
        {konevcny2015federated}
\bibfield{author}{\bibinfo{person}{Jakub Kone{\v{c}}n{\`y}},
  \bibinfo{person}{Brendan McMahan}, {and} \bibinfo{person}{Daniel Ramage}.}
  \bibinfo{year}{2015}\natexlab{}.
\newblock \showarticletitle{Federated optimization: Distributed optimization
  beyond the datacenter}.
\newblock \bibinfo{journal}{\emph{arXiv preprint arXiv:1511.03575}}
  (\bibinfo{year}{2015}).
\newblock


\bibitem[\protect\citeauthoryear{Kraska, Beutel, Chi, Dean, and
  Polyzotis}{Kraska et~al\mbox{.}}{2018}]%
        {kraska2018case}
\bibfield{author}{\bibinfo{person}{Tim Kraska}, \bibinfo{person}{Alex Beutel},
  \bibinfo{person}{Ed~H Chi}, \bibinfo{person}{Jeffrey Dean}, {and}
  \bibinfo{person}{Neoklis Polyzotis}.} \bibinfo{year}{2018}\natexlab{}.
\newblock \showarticletitle{The case for learned index structures}. In
  \bibinfo{booktitle}{\emph{Proceedings of the 2018 International Conference on
  Management of Data}}. \bibinfo{pages}{489--504}.
\newblock


\bibitem[\protect\citeauthoryear{Leis, Gubichev, Mirchev, Boncz, Kemper, and
  Neumann}{Leis et~al\mbox{.}}{2015}]%
        {howgoodare}
\bibfield{author}{\bibinfo{person}{Viktor Leis}, \bibinfo{person}{Andrey
  Gubichev}, \bibinfo{person}{Atanas Mirchev}, \bibinfo{person}{Peter Boncz},
  \bibinfo{person}{Alfons Kemper}, {and} \bibinfo{person}{Thomas Neumann}.}
  \bibinfo{year}{2015}\natexlab{}.
\newblock \showarticletitle{How Good Are Query Optimizers, Really?}
\newblock \bibinfo{journal}{\emph{Proc. VLDB Endow}}  \bibinfo{volume}{9}
  (\bibinfo{year}{2015}), \bibinfo{pages}{204--215}.
\newblock


\bibitem[\protect\citeauthoryear{Li, Zhou, Li, and Gao}{Li
  et~al\mbox{.}}{2019}]%
        {li2019qtune}
\bibfield{author}{\bibinfo{person}{Guoliang Li}, \bibinfo{person}{Xuanhe Zhou},
  \bibinfo{person}{Shifu Li}, {and} \bibinfo{person}{Bo Gao}.}
  \bibinfo{year}{2019}\natexlab{}.
\newblock \showarticletitle{Qtune: A query-aware database tuning system with
  deep reinforcement learning}.
\newblock \bibinfo{journal}{\emph{Proceedings of the VLDB Endowment}}
  \bibinfo{volume}{12}, \bibinfo{number}{12} (\bibinfo{year}{2019}),
  \bibinfo{pages}{2118--2130}.
\newblock


\bibitem[\protect\citeauthoryear{Ma, Ding, Das, and Swaminathan}{Ma
  et~al\mbox{.}}{2020}]%
        {ma2020active}
\bibfield{author}{\bibinfo{person}{Lin Ma}, \bibinfo{person}{Bailu Ding},
  \bibinfo{person}{Sudipto Das}, {and} \bibinfo{person}{Adith Swaminathan}.}
  \bibinfo{year}{2020}\natexlab{}.
\newblock \showarticletitle{Active learning for ML enhanced database systems}.
  In \bibinfo{booktitle}{\emph{Proceedings of the 2020 ACM SIGMOD International
  Conference on Management of Data}}. \bibinfo{pages}{175--191}.
\newblock


\bibitem[\protect\citeauthoryear{Marcus, Negi, Mao, Zhang, Alizadeh, Kraska,
  Papaemmanouil, and Tatbul}{Marcus et~al\mbox{.}}{2019}]%
        {marcus2019neo}
\bibfield{author}{\bibinfo{person}{Ryan Marcus}, \bibinfo{person}{Parimarjan
  Negi}, \bibinfo{person}{Hongzi Mao}, \bibinfo{person}{Chi Zhang},
  \bibinfo{person}{Mohammad Alizadeh}, \bibinfo{person}{Tim Kraska},
  \bibinfo{person}{Olga Papaemmanouil}, {and} \bibinfo{person}{Nesime Tatbul}.}
  \bibinfo{year}{2019}\natexlab{}.
\newblock \showarticletitle{Neo: A learned query optimizer}.
\newblock \bibinfo{journal}{\emph{PVLDB}} (\bibinfo{year}{2019}).
\newblock


\bibitem[\protect\citeauthoryear{Marcus and Papaemmanouil}{Marcus and
  Papaemmanouil}{[n.d.]}]%
        {marcus2016wisedb}
\bibfield{author}{\bibinfo{person}{Ryan Marcus} {and} \bibinfo{person}{Olga
  Papaemmanouil}.} \bibinfo{year}{[n.d.]}\natexlab{}.
\newblock \showarticletitle{Wisedb: A learning-based workload management
  advisor for cloud databases}.
\newblock \bibinfo{journal}{\emph{PVLDB}} (\bibinfo{year}{[n.\,d.]}),
  \bibinfo{pages}{780--791}.
\newblock


\bibitem[\protect\citeauthoryear{Marcus and Papaemmanouil}{Marcus and
  Papaemmanouil}{2018}]%
        {marcus2018deep}
\bibfield{author}{\bibinfo{person}{Ryan Marcus} {and} \bibinfo{person}{Olga
  Papaemmanouil}.} \bibinfo{year}{2018}\natexlab{}.
\newblock \showarticletitle{Deep reinforcement learning for join order
  enumeration}. In \bibinfo{booktitle}{\emph{Proceedings of the First
  International Workshop on Exploiting Artificial Intelligence Techniques for
  Data Management}}. \bibinfo{pages}{1--4}.
\newblock


\bibitem[\protect\citeauthoryear{Marcus and Papaemmanouil}{Marcus and
  Papaemmanouil}{2019}]%
        {marcus2019plan}
\bibfield{author}{\bibinfo{person}{Ryan Marcus} {and} \bibinfo{person}{Olga
  Papaemmanouil}.} \bibinfo{year}{2019}\natexlab{}.
\newblock \showarticletitle{Plan-structured deep neural network models for
  query performance prediction}.
\newblock \bibinfo{journal}{\emph{arXiv preprint arXiv:1902.00132}}
  (\bibinfo{year}{2019}).
\newblock


\bibitem[\protect\citeauthoryear{Nathan, Ding, Alizadeh, and Kraska}{Nathan
  et~al\mbox{.}}{2020}]%
        {nathan2020learning}
\bibfield{author}{\bibinfo{person}{Vikram Nathan}, \bibinfo{person}{Jialin
  Ding}, \bibinfo{person}{Mohammad Alizadeh}, {and} \bibinfo{person}{Tim
  Kraska}.} \bibinfo{year}{2020}\natexlab{}.
\newblock \showarticletitle{Learning multi-dimensional indexes}. In
  \bibinfo{booktitle}{\emph{Proceedings of the 2020 ACM SIGMOD International
  Conference on Management of Data}}. \bibinfo{pages}{985--1000}.
\newblock


\bibitem[\protect\citeauthoryear{Negi, Marcus, Kipf, Mao, Tatbul, Kraska, and
  Alizadeh}{Negi et~al\mbox{.}}{2021}]%
        {negi2021flow}
\bibfield{author}{\bibinfo{person}{Parimarjan Negi}, \bibinfo{person}{Ryan
  Marcus}, \bibinfo{person}{Andreas Kipf}, \bibinfo{person}{Hongzi Mao},
  \bibinfo{person}{Nesime Tatbul}, \bibinfo{person}{Tim Kraska}, {and}
  \bibinfo{person}{Mohammad Alizadeh}.} \bibinfo{year}{2021}\natexlab{}.
\newblock \showarticletitle{Flow-Loss: Learning Cardinality Estimates That
  Matter}.
\newblock \bibinfo{journal}{\emph{arXiv preprint arXiv:2101.04964}}
  (\bibinfo{year}{2021}).
\newblock


\bibitem[\protect\citeauthoryear{Ortiz, Balazinska, Gehrke, and Keerthi}{Ortiz
  et~al\mbox{.}}{2018}]%
        {ortiz2018learning}
\bibfield{author}{\bibinfo{person}{Jennifer Ortiz}, \bibinfo{person}{Magdalena
  Balazinska}, \bibinfo{person}{Johannes Gehrke}, {and}
  \bibinfo{person}{S~Sathiya Keerthi}.} \bibinfo{year}{2018}\natexlab{}.
\newblock \showarticletitle{Learning state representations for query
  optimization with deep reinforcement learning}. In
  \bibinfo{booktitle}{\emph{Proceedings of the Second Workshop on Data
  Management for End-To-End Machine Learning}}. \bibinfo{pages}{1--4}.
\newblock


\bibitem[\protect\citeauthoryear{Ranzato, Chopra, Auli, and Zaremba}{Ranzato
  et~al\mbox{.}}{2015}]%
        {ranzato2015sequence}
\bibfield{author}{\bibinfo{person}{Marc'Aurelio Ranzato},
  \bibinfo{person}{Sumit Chopra}, \bibinfo{person}{Michael Auli}, {and}
  \bibinfo{person}{Wojciech Zaremba}.} \bibinfo{year}{2015}\natexlab{}.
\newblock \showarticletitle{Sequence level training with recurrent neural
  networks}.
\newblock \bibinfo{journal}{\emph{arXiv preprint arXiv:1511.06732}}
  (\bibinfo{year}{2015}).
\newblock


\bibitem[\protect\citeauthoryear{Sheng, Tomasic, Zhang, and Pavlo}{Sheng
  et~al\mbox{.}}{2019}]%
        {sheng2019scheduling}
\bibfield{author}{\bibinfo{person}{Yangjun Sheng}, \bibinfo{person}{Anthony
  Tomasic}, \bibinfo{person}{Tieying Zhang}, {and} \bibinfo{person}{Andrew
  Pavlo}.} \bibinfo{year}{2019}\natexlab{}.
\newblock \showarticletitle{Scheduling OLTP transactions via learned abort
  prediction}. In \bibinfo{booktitle}{\emph{Proceedings of the Second
  International Workshop on Exploiting Artificial Intelligence Techniques for
  Data Management}}. \bibinfo{pages}{1--8}.
\newblock


\bibitem[\protect\citeauthoryear{Shiv and Quirk}{Shiv and Quirk}{2019}]%
        {shiv2019novel}
\bibfield{author}{\bibinfo{person}{Vighnesh Shiv} {and} \bibinfo{person}{Chris
  Quirk}.} \bibinfo{year}{2019}\natexlab{}.
\newblock \showarticletitle{Novel positional encodings to enable tree-based
  transformers}.
\newblock \bibinfo{journal}{\emph{Advances in Neural Information Processing
  Systems}}  \bibinfo{volume}{32} (\bibinfo{year}{2019}),
  \bibinfo{pages}{12081--12091}.
\newblock


\bibitem[\protect\citeauthoryear{Siddiqui, Jindal, Qiao, Patel, and
  Le}{Siddiqui et~al\mbox{.}}{2020}]%
        {siddiqui2020cost}
\bibfield{author}{\bibinfo{person}{Tarique Siddiqui}, \bibinfo{person}{Alekh
  Jindal}, \bibinfo{person}{Shi Qiao}, \bibinfo{person}{Hiren Patel}, {and}
  \bibinfo{person}{Wangchao Le}.} \bibinfo{year}{2020}\natexlab{}.
\newblock \showarticletitle{Cost models for big data query processing:
  Learning, retrofitting, and our findings}. In
  \bibinfo{booktitle}{\emph{Proceedings of the 2020 ACM SIGMOD International
  Conference on Management of Data}}. \bibinfo{pages}{99--113}.
\newblock


\bibitem[\protect\citeauthoryear{Sun and Li}{Sun and Li}{2019}]%
        {sun2019end}
\bibfield{author}{\bibinfo{person}{Ji Sun} {and} \bibinfo{person}{Guoliang
  Li}.} \bibinfo{year}{2019}\natexlab{}.
\newblock \showarticletitle{An end-to-end learning-based cost estimator}.
\newblock \bibinfo{journal}{\emph{arXiv preprint arXiv:1906.02560}}
  (\bibinfo{year}{2019}).
\newblock


\bibitem[\protect\citeauthoryear{Tang, Fan, Li, Tu, Du, Li, Madden, and
  Ouzzani}{Tang et~al\mbox{.}}{2020}]%
        {tang2020relational}
\bibfield{author}{\bibinfo{person}{Nan Tang}, \bibinfo{person}{Ju Fan},
  \bibinfo{person}{Fangyi Li}, \bibinfo{person}{Jianhong Tu},
  \bibinfo{person}{Xiaoyong Du}, \bibinfo{person}{Guoliang Li},
  \bibinfo{person}{Sam Madden}, {and} \bibinfo{person}{Mourad Ouzzani}.}
  \bibinfo{year}{2020}\natexlab{}.
\newblock \showarticletitle{Relational Pretrained Transformers towards
  Democratizing Data Preparation [Vision]}.
\newblock \bibinfo{journal}{\emph{arXiv preprint arXiv:2012.02469}}
  (\bibinfo{year}{2020}).
\newblock


\bibitem[\protect\citeauthoryear{Trummer}{Trummer}{2019}]%
        {trummer2019exact}
\bibfield{author}{\bibinfo{person}{Immanuel Trummer}.}
  \bibinfo{year}{2019}\natexlab{}.
\newblock \showarticletitle{Exact cardinality query optimization with bounded
  execution cost}. In \bibinfo{booktitle}{\emph{Proceedings of the 2019
  International Conference on Management of Data}}. \bibinfo{pages}{2--17}.
\newblock


\bibitem[\protect\citeauthoryear{Vaswani, Shazeer, Parmar, Uszkoreit, Jones,
  Gomez, Kaiser, and Polosukhin}{Vaswani et~al\mbox{.}}{2017}]%
        {transformer}
\bibfield{author}{\bibinfo{person}{Ashish Vaswani}, \bibinfo{person}{Noam
  Shazeer}, \bibinfo{person}{Niki Parmar}, \bibinfo{person}{Jakob Uszkoreit},
  \bibinfo{person}{Llion Jones}, \bibinfo{person}{Aidan~N. Gomez},
  \bibinfo{person}{Lukasz Kaiser}, {and} \bibinfo{person}{Illia Polosukhin}.}
  \bibinfo{year}{2017}\natexlab{}.
\newblock \showarticletitle{Attention is all you need}.
\newblock \bibinfo{journal}{\emph{In Advances in neural information processing
  systems}} (\bibinfo{year}{2017}).
\newblock


\bibitem[\protect\citeauthoryear{Wang, Qu, Wu, Wang, and Zhou}{Wang
  et~al\mbox{.}}{2020}]%
        {wang2020ready}
\bibfield{author}{\bibinfo{person}{Xiaoying Wang}, \bibinfo{person}{Changbo
  Qu}, \bibinfo{person}{Weiyuan Wu}, \bibinfo{person}{Jiannan Wang}, {and}
  \bibinfo{person}{Qingqing Zhou}.} \bibinfo{year}{2020}\natexlab{}.
\newblock \bibinfo{title}{Are We Ready For Learned Cardinality Estimation?}
\newblock
\newblock
\showeprint[arxiv]{2012.06743}~[cs.DB]


\bibitem[\protect\citeauthoryear{Wu and Shaikhha}{Wu and Shaikhha}{2020}]%
        {wu2020bayescard}
\bibfield{author}{\bibinfo{person}{Ziniu Wu} {and} \bibinfo{person}{Amir
  Shaikhha}.} \bibinfo{year}{2020}\natexlab{}.
\newblock \showarticletitle{BayesCard: A Unified Bayesian Framework for
  Cardinality Estimation}.
\newblock \bibinfo{journal}{\emph{arXiv preprint arXiv:2012.14743}}
  (\bibinfo{year}{2020}).
\newblock


\bibitem[\protect\citeauthoryear{Yang, Kamsetty, Luan, Liang, Duan, Chen, and
  Stoica}{Yang et~al\mbox{.}}{2020}]%
        {NeuroCard}
\bibfield{author}{\bibinfo{person}{Zongheng Yang}, \bibinfo{person}{Amog
  Kamsetty}, \bibinfo{person}{Sifei Luan}, \bibinfo{person}{Eric Liang},
  \bibinfo{person}{Yan Duan}, \bibinfo{person}{Xi Chen}, {and}
  \bibinfo{person}{Ion Stoica}.} \bibinfo{year}{2020}\natexlab{}.
\newblock \showarticletitle{NeuroCard: One Cardinality Estimator for All
  Tables}.
\newblock \bibinfo{journal}{\emph{arxiv}} (\bibinfo{year}{2020}).
\newblock


\bibitem[\protect\citeauthoryear{Yang, Liang, Kamsetty, Wu, Duan, Chen, Abbeel,
  Hellerstein, Krishnan, and Stoica}{Yang et~al\mbox{.}}{2019}]%
        {naru}
\bibfield{author}{\bibinfo{person}{Zongheng Yang}, \bibinfo{person}{Eric
  Liang}, \bibinfo{person}{Amog Kamsetty}, \bibinfo{person}{Chenggang Wu},
  \bibinfo{person}{Yan Duan}, \bibinfo{person}{Xi Chen},
  \bibinfo{person}{Pieter Abbeel}, \bibinfo{person}{Joseph~M. Hellerstein},
  \bibinfo{person}{Sanjay Krishnan}, {and} \bibinfo{person}{Ion Stoica}.}
  \bibinfo{year}{2019}\natexlab{}.
\newblock \showarticletitle{Deep unsupervised cardinality estimation}.
\newblock \bibinfo{journal}{\emph{Proceedings of the VLDB Endowment}}
  (\bibinfo{year}{2019}).
\newblock


\bibitem[\protect\citeauthoryear{Yu, Li, Chai, and Tang}{Yu
  et~al\mbox{.}}{2020}]%
        {yu2020reinforcement}
\bibfield{author}{\bibinfo{person}{Xiang Yu}, \bibinfo{person}{Guoliang Li},
  \bibinfo{person}{Chengliang Chai}, {and} \bibinfo{person}{Nan Tang}.}
  \bibinfo{year}{2020}\natexlab{}.
\newblock \showarticletitle{Reinforcement learning with tree-lstm for join
  order selection}. In \bibinfo{booktitle}{\emph{2020 IEEE 36th International
  Conference on Data Engineering (ICDE)}}. IEEE, \bibinfo{pages}{1297--1308}.
\newblock


\bibitem[\protect\citeauthoryear{Zhang, Liu, Zhou, Li, Xiao, Cheng, Xing, Wang,
  Cheng, Liu, et~al\mbox{.}}{Zhang et~al\mbox{.}}{2019}]%
        {zhang2019end}
\bibfield{author}{\bibinfo{person}{Ji Zhang}, \bibinfo{person}{Yu Liu},
  \bibinfo{person}{Ke Zhou}, \bibinfo{person}{Guoliang Li},
  \bibinfo{person}{Zhili Xiao}, \bibinfo{person}{Bin Cheng},
  \bibinfo{person}{Jiashu Xing}, \bibinfo{person}{Yangtao Wang},
  \bibinfo{person}{Tianheng Cheng}, \bibinfo{person}{Li Liu}, {et~al\mbox{.}}}
  \bibinfo{year}{2019}\natexlab{}.
\newblock \showarticletitle{An end-to-end automatic cloud database tuning
  system using deep reinforcement learning}. In
  \bibinfo{booktitle}{\emph{Proceedings of the 2019 International Conference on
  Management of Data}}. \bibinfo{pages}{415--432}.
\newblock


\bibitem[\protect\citeauthoryear{Zhou, Chai, Li, and Sun}{Zhou
  et~al\mbox{.}}{2020}]%
        {zhou2020database}
\bibfield{author}{\bibinfo{person}{Xuanhe Zhou}, \bibinfo{person}{Chengliang
  Chai}, \bibinfo{person}{Guoliang Li}, {and} \bibinfo{person}{Ji Sun}.}
  \bibinfo{year}{2020}\natexlab{}.
\newblock \showarticletitle{Database meets artificial intelligence: A survey}.
\newblock \bibinfo{journal}{\emph{IEEE Transactions on Knowledge and Data
  Engineering}} (\bibinfo{year}{2020}).
\newblock


\bibitem[\protect\citeauthoryear{Zhu, Wu, Han, Zeng, Pfadler, Qian, Zhou, and
  Cui}{Zhu et~al\mbox{.}}{2021}]%
        {zhu2020flat}
\bibfield{author}{\bibinfo{person}{Rong Zhu}, \bibinfo{person}{Ziniu Wu},
  \bibinfo{person}{Yuxing Han}, \bibinfo{person}{Kai Zeng},
  \bibinfo{person}{Andreas Pfadler}, \bibinfo{person}{Zhengping Qian},
  \bibinfo{person}{Jingren Zhou}, {and} \bibinfo{person}{Bin Cui}.}
  \bibinfo{year}{2021}\natexlab{}.
\newblock \showarticletitle{FLAT: Fast, Lightweight and Accurate Method for
  Cardinality Estimation}.
\newblock \bibinfo{journal}{\emph{VLDB}} (\bibinfo{year}{2021}).
\newblock


\end{thebibliography}
	
\end{document}